\documentclass[11pt,a4paper]{article}
\pdfoutput=1
\usepackage{jheppub}

\usepackage{cite}

\usepackage{multirow, graphicx,amssymb,url,mathrsfs,amsmath}
\usepackage{wrapfig,boxedminipage,subfigure,epsfig}
\usepackage{amsxtra,amstext,latexsym,dsfont,amsfonts}
\usepackage{color}
\usepackage[dvipsnames]{xcolor}
\usepackage{float}
\usepackage{slashed}
\usepackage{calligra}
\DeclareFontShape{T1}{calligra}{m}{n}{<->s*[2.2]callig15}{}
\DeclareMathAlphabet{\mathcalligra}{T1}{calligra}{m}{n}

\newcommand{\be}{\begin{equation}}
\newcommand{\ee}{\end{equation}}
\newcommand{\bea}{\begin{eqnarray}}
\newcommand{\eea}{\end{eqnarray}}

\newcommand{\mt}[1]{\textrm{\tiny #1}}

\newcommand{\dd}{\mathrm{d}}

\renewcommand{\Im}{{ \rm Im}}

\newcommand{\nr}{{\mathcalligra{r}}}

\newcommand{\kyr}[1]{{\color{red}{#1}}}
\newcommand{\kyb}[1]{{\color{blue}{#1}}}
\newcommand{\vj}[1]{{\color{purple}{#1}}}
\newcommand{\ks}[1]{{\color{teal}{#1}}}
\newcommand{\yj}[1]{{\color{brown}{#1}}}

\title{
Classifying pole-skipping points}

\author[a]{Yongjun Ahn,}
\author[a]{Viktor Jahnke,}
\author[b]{Hyun-Sik Jeong,}
\author[a]{Keun-Young Kim,}
\author[a]{Kyung-Sun Lee,}
\author[a]{and Mitsuhiro Nishida}

\affiliation[a]{School of Physics and Chemistry, Gwangju Institute of Science and Technology, 123 Cheomdan-gwagiro, Gwangju 61005, Korea}
\affiliation[b]{Center for Quantum Spacetime, Sogang University, Seoul 04107, Korea}

\emailAdd{yongjunahn619@gmail.com}
\emailAdd{viktorjahnke@gist.ac.kr}
\emailAdd{hyunsiksgu@sogang.ac.kr}
\emailAdd{fortoe@gist.ac.kr}
\emailAdd{kyungsun.cogito.lee@gmail.com }
\emailAdd{mnishida@gist.ac.kr}

\abstract{
We clarify general mathematical and physical properties of pole-skipping points. For this purpose, we analyse scalar and vector fields in hyperbolic space. This setup is chosen because it is simple enough to allow us to obtain analytical expressions for the Green's function and check everything explicitly, while it contains all the essential features of pole-skipping points. We classify pole-skipping points in three types (type-I, II, III). Type-I and Type-II are distinguished by the (limiting) behavior of the Green's function near the 
pole-skipping points. Type-III can arise at non-integer $i\omega$ values, which is due to a specific UV condition, contrary to the types I and II, which are related to a non-unique near horizon boundary condition. We also clarify the relation between the pole-skipping structure of the Green's function and the near horizon analysis. We point out that there are subtle cases  where the near horizon analysis alone may not be able to capture the existence and properties of the pole-skipping points. 
 }

\begin{document}
\maketitle



\section{Introduction}

In recent years, the so-called pole-skipping phenomenon~\cite{Grozdanov_2018,Blake_2018,Blake:2018leo} has attracted much attention, specially because of its connection with quantum chaos.\footnote{For a review on quantum chaos, in the context of holography,  we refer to~\cite{Jahnke:2018off, Jahnke:2019gxr} and references therein.}  Pole-skipping points are special values of frequency and wave number at which a momentum space Green's function is not well defined. This phenomenon is particularly important in the context of gauge-gravity duality \cite{Maldacena:1997re, Witten:1998qj,Gubser:1998bc}, because it reveals properties of the boundary theory that are completely encoded in the region near the black hole horizon.

According to the holographic dictionary, boundary correlators are related to solutions of the classical equations of motion in the bulk. In particular, a retarded Green's function in momentum space can be read off from the near boundary behavior of perturbations satisfying ingoing boundary conditions at the black hole horizon. In general, for generic values of frequency and wave number, $(\omega, k)$, the ingoing solution is unique, which leads to the uniqueness of the Green's function. 

There are, however, some special values of frequency and wave number $(\omega_*, k_*)$ at which the ingoing solution is not unique. As a consequence, the Green's function is ill-defined at those points \cite{Blake:2018leo}. These special points are known as {\it pole-skipping points}. They were first discovered in energy density retarded two-point functions, where they were shown to be related to behavior of out-of-time-order correlators (OTOCs)~\cite{larkin1969quasiclassical, Kitaev-2014, Maldacena:2015waa}\footnote{OTOC in the context of quantum many body systems and quantum mechanics we refer to~\cite{Huh:2020joh, Hashimoto:2020xfr} and references therein.} that characterize chaotic behavior~\cite{Grozdanov_2018}.  More specifically,
\begin{equation} 
    \omega_* = i \lambda_L\,,\,\,\,\, k= i \frac{\lambda_L}{v_B}
\end{equation}
where $\lambda_L =2 \pi T$ is the Lyapunov exponent, and $v_B$ is the butterfly velocity, which control the exponential behavior of OTOCs, i.e., $ \langle V(0) W(t,x) V(0) W(t,x) \rangle \sim c_0+c_1 e^{ \lambda_L (t-|x|/v_B)}$, where $c_0$ and $c_1$ are constants. The relation between pole-skipping and the non-uniqueness of the incoming solution was discovered in \cite{Blake:2018leo}, where it was also shown that a near horizon analysis was enough to compute the pole-skipping point. It was later realized that pole-skipping is a much more general phenomenon, that takes place for correlators involving other component of the stress energy tensor, and also for vector and scalar fields, although in those cases the pole-skipping points are not in general related to chaos \cite{Blake:2019otz,Grozdanov:2019uhi}. Moreover, \cite{Blake:2019otz,Grozdanov:2019uhi} revealed the existence of a infinite set of pole-skipping points, occurring at higher Matsubara frequencies. The existence of pole-skipping for fermions was confirmed in \cite{_eplak_2020}. Other recent development involving pole-skipping include for instance \cite{Natsuume:2019sfp,Natsuume:2019xcy,Natsuume:2019vcv,Grozdanov_2019,Grozdanov:2020koi,Li:2019bgc,   Liu:2020yaf, Abbasi:2020ykq, Jansen:2020hfd,  Wu:2019esr, Abbasi:2019rhy, Haehl:2018izb, Das:2019tga, Ramirez:2020qer}.

In all the above examples, pole-skipping was studied for boundary theories living in a flat spacetime, where one can decompose the bulk perturbations and the boundary Green's function in terms of plane waves. Pole-skipping also takes place in hyperbolic space \cite{Ahn:2019rnq, Haehl:2019eae}. In this case one has to decompose the bulk field perturbation and the boundary Green's function in terms of scalar harmonics of the Laplacian operator in hyperbolic space. The use of hyperbolic space is convenient because it allows us to obtain analytic results in both sides of the AdS/CFT duality. In particular, by considering a Rindler-AdS geometry, the boundary description is given in terms of a CFT in hyperbolic space, while the bulk geometry is just the Rindler wedge of AdS. In this holographic set up, one can obtain the exact Green's function of scalar and vector fields, and compute the full set of pole-skipping points in both sides of the AdS/CFT duality \cite{Ahn:2020bks}.

The authors of \cite{Ahn:2020bks} confirmed that the leading pole-skipping points of scalar and vector fields in a Rindler-AdS geometry can also be obtained by a near horizon analysis, but they did not check whether the full tower of pole-skipping points can also be obtained in this way. In this work, we study the full tower of pole-skipping points of scalar and vector fields in a Rindler-AdS geometry using a near horizon analysis, following the formalism developed in \cite{Blake:2019otz}. 

Analysing the results, we find that, for certain values of the scaling dimension $\Delta$ and the spacetime dimension $d$, there appears to be some anomalous points \cite{Blake:2019otz}, whose properties have not been fully clarified yet. Near the anomalous points, the Green's function does not depend on the ratio of the deviation ($\delta \omega/\delta k$) from the pole-skipping points. In some cases, the anomalous pole-skipping point looked like an intersection of two lines of poles\footnote{It turns out to be not the case.}. 
Building on the analysis in \cite{Blake:2019otz} and \cite{Natsuume:2019vcv} and 
based on our concrete computation from field theory and holography, we clarify the nature of the anomalous pole-skipping points. We call it type-II pole-skipping point to distinguish it from the usual one, which we call type-I pole-skipping point. Furthermore, we find a possibility that there is another type of the pole-skipping point with non-integer $i\omega$ values, which we call type-III.


Let us highlight some of the main properties of the different types of pole-skipping points:

\begin{itemize}
    \item type I and type II: in theses cases, $i \omega$ takes integer values, and the non-uniqueness of the Green's function $\mathcal{G}^{R}(\omega, k)$ at the pole-skipping point is related to the non-uniqueness of incoming solution at the black holes horizon. This basically means that the value that $\mathcal{G}^{R}(\omega, k)$ takes at the pole-skipping point $(\omega_*,k_*)$ depends on the particular path $\omega(k)$ we take to approach this point. In particular, we can always choose a path in which $\mathcal{G}^{R}(\omega, k)$ is constant. For type I points, the Green's function is constant along straight lines that pass though the pole-skipping points, while for type II points, the Green's function is constant along quadratic or higher order curves passing through the pole-skipping point. 
    Type I points occur when a line of zeros intersects a line of poles, or, more generally, when the same number of zero-lines and pole-lines intersect at a point. For type II points, a different number of zero-lines and pole-lines intersect at a point.
    
    \item type III: in this case $i \omega$ takes in general non-integer values, and the non-uniqueness of the Green's function at the pole-skipping point is not related to the incoming solution at the horizon, which is unique in this case. Actually, for type III points, the non-uniqueness results from a UV property of the theory, rather than a IR property, i.e., it is not related to the black hole horizon.
\end{itemize}


We stress that the above classification of poles-skipping points was facilitated because we have analytic control of the retarded Green's function of scalar and vector operators in hyperbolic space. Since the knowledge of the Green's function is not always available, it is important to know whether this classification of pole-skipping points can also be done using a near horizon analysis, which is easily applicable in much more general geometries. In the main text, we show that this classification from the near horizon analysis is possible.  We point out that there is a subtlety: sometimes the near horizon analysis alone is not enough to capture the existence and properties of the pole-skipping points.  

 This paper is organized as follows. In Sec.\ref{SNH}, we study the pole-skipping points of scalar and vector fields in hyperbolic space using a near horizon analysis and compare the results with the ones obtained using field theory. 
 In Sec.\ref{section3},
 we classify pole-skipping points in two types (type-I and type-II). After showing a simple example for type-II pole-skipping points, we develop a general argument for the type-I and type-II pole-skipping points from the Green's function perspective.
 In Sec.\ref{Sec4}, we investigate how to identify the type-I and type-II pole-skipping points from  a near horizon analysis.
 A subtlety in the near horizon analysis is discussed in 
 Sec.\ref{Sec5}. Based on the observation from this subtlety, we find another type of the pole-skipping point in the near horizon analysis, type-III, and discuss its mechanism.
 We conclude in Sec.\ref{discussion}.
 
\section{Pole-skipping points from near horizon analysis and field theory\label{SNH}}
Let us recall that a pole-skipping point is defined as a point in momentum space at which the Green's function is ill-defined (e.g., it takes the form of $G^R\sim\frac{0}{0}$) \cite{Blake:2018leo, Blake:2019otz, Natsuume:2019sfp, Natsuume:2019vcv, Natsuume:2019xcy}. From the holographic perspective, one  possibility for pole-skipping points is the ambiguity of boundary conditions at the horizon. If there is no unique way to impose incoming boundary conditions on the horizon at certain frequencies and wave numbers, the Green's function will not be unique at those points. Such points can be found by analyzing the bulk equation of motion near the black hole horizon.

In this section, we build `towers' of pole-skipping points \cite{Blake:2019otz} of scalar and massless vector fields by analyzing the bulk equations of motion near the black hole horizon. In the case in which the boundary theory lives in flat space, the near horizon analysis for higher (Matsubara) frequencies has been done in \cite{Blake:2019otz, Natsuume:2019vcv, Abbasi:2019rhy}. Following the same method, we consider  the case where the boundary theory lives in hyperbolic space. For simplicity we ignore back-reactions.

The pole-skipping points for scalar and vector fields in hyperbolic space have been analytically obtained from a field theory computation~\cite{Ahn:2020bks}. We confirm that our holographic near horizon analysis completely agrees with the field theory result.

\paragraph{Setup: geometry and coordinate systems}
As a classical solution of the Einstein-Hilbert action
\begin{equation}
\label{EHaction}
S = \int d^{d+1} x \sqrt{-g}\left[R+\frac{d\left(d-1\right)}{ \ell_{\mathrm{AdS}}^2} \right]\,,
\end{equation}
we consider the Rindler-AdS$_{d+1}$ geometry 
\begin{equation}
\label{AdSRind}
\begin{split}
ds^2&=\frac{\ell_{\mathrm{AdS}}^2}{z^2}\left(-f(z)dt^2+\frac{dz^2}{f(z)} +\ell_{\mathrm{AdS}}^2 \left(d\chi^2+\sinh^2\chi \,d\Omega_{d-2}^2 \right)\right)\,,\\
f(z)&=1-\frac{z^2}{\ell_{\mathrm{AdS}}^2}\,,
\end{split}
\end{equation}
where $\ell_{\mathrm{AdS}}$ is an $\mathrm{AdS}$ radius and we consider that the black hole horizon radius is $\ell_{\mathrm{AdS}}$. The Hawking temperature is
\begin{align}
\label{HawkingT}
T=\frac{1}{2\pi \ell_{\mathrm{AdS}}}\,.
\end{align}
Although the analysis can be done for a general hyperbolic black hole with any  horizon radius apart from $\ell_{\mathrm{AdS}}$, we will focus on \eqref{AdSRind} because, in this case, the field theory dual analysis can be done analytically so we may be able to compare our results  with the field theory results. 

In the following, we set $\ell_{\mathrm{AdS}}=1$. For the near horizon analysis, it is useful to introduce the incoming Eddington-Finkelstein coordinate $v$ as
\begin{equation}
\label{inEFco}
v=t-z_*\,,\,\,\,\,\,\,dz_*=\frac{dz}{f}\,,
\end{equation}
in terms of which the metric \eqref{AdSRind} becomes
\begin{equation}
\begin{split}
\dd s^2&=-\frac{f(z)}{z^2}\dd v^2-\frac{2}{z^2} \dd v \dd z+\frac{1}{z^2}\left( \dd\chi^2+\sinh^2\chi \dd\Omega_{d-2}^2 \right)\,, \\
f(z)&=1-z^2
\,.
\end{split}\label{inEFhypBH} 
\end{equation}

\subsection{Scalar field} \label{scalar}

\subsubsection{Near horizon analysis} \label{scalar11}
In this section, we compute the full set of pole-skipping points of a massive scalar field by analysing the near horizon equations of motion. We assume that the action of the scalar field has the form
\be
S_\mt{scalar}=-\frac{1}{2}\int d^{d+1}x \sqrt{-g} \left(g^{\mu \nu} \partial_{\mu} \varphi \partial_{\nu} \varphi+m^2 \varphi^2 \right)\label{saction}\,,
\ee
where the background geometry is \eqref{AdSRind}. In  incoming Eddington-Finkelstein coordinates \eqref{inEFco}, the equation of motion for the massive scalar field can be written as
\begin{equation}
\label{KGeq}
z^{d+1} \partial_z \left(z^{1-d} f(z) \partial_z \varphi \right)+z(d-1) \partial_{v} \varphi
-2z^2 \partial_v \partial_z \varphi+z^2 \square_{H_{d-1}}\varphi -m^2 \varphi=0\,, 
\end{equation}
where 
\begin{equation}
\square_{H^{d-1}}= \partial_{\chi}^2+(d-2) \coth \chi \partial_{\chi}+\frac{1}{\sinh^2\chi} \square_{S^{d-2}} \,,
\end{equation}
is the Laplacian operator in $\mathbb{H}^{d-1}$.

Let us consider the `Fourier' transformation
\begin{equation}
\label{FTscal}
\varphi(v,z,\chi)= \int \dd \omega \, \dd k\, \phi(z;\omega, k) e^{-i \omega v} \mathbb{S}_{k}(\chi)\,,
\end{equation}
where $\mathbb{S}_{k}(\chi)$ is an eigenfunction of $\square_{H^{d-1}}$, which is defined as
\begin{equation}
\label{EVeq1}
\left(\square_{H_{d-1}}+{k}^2+\left(\frac{d-2}{2}\right)^2\right)\mathbb{S}_{k}(\chi)=0.
\end{equation}
In terms of the Fourier mode, $\phi(z;\omega, k)$, the equation of motion \eqref{KGeq} boils down to
\begin{equation} \label{KGeq2}
\begin{split}
&\phi ''(z)-\frac{\mathcal{F}_1(d,\omega,z)}{z f(z)}\phi'(z)-\frac{\mathcal{F}_2(d,\omega,z,k)}{z^2 f(z)}\phi (z)=0\,,\\
&\qquad \mathcal{F}_1(d,\omega,z)=(d-1) f(z)-z \left(f'(z)+2 i \omega \right)\,,\\
&\qquad \mathcal{F}_2(d,\omega,k,z)= \left(k^2+\left(\frac{d-2}{2}\right)^2\right)z^2+i (d-1) \omega  z+m^2\,,
\end{split}
\end{equation}
where we omit the $k$ and $\omega$ dependence in $\phi(z;\omega,k)$ to avoid clutter. 

The equation \eqref{KGeq2} is a second-order differential equation with a regular singular point at the horizon, at $z=1$. By solving the indicial equation near the singular point, we obtain two independent solutions
\begin{equation} \label{gscgs}
    \phi(z)\ \sim \ (1-z)^0  \qquad  \mathrm{or} \qquad (1-z)^{i\omega} \,.
\end{equation}
Usually, the second solution is discarded since it is not regular. The resulting solution $\phi(z)\ \sim \ (1-z)^0$ is unique and amounts to incoming boundary condition at the horizon. Thanks to this condition, the holographic Green's function is uniquely determined. 

However, if the frequency is given by $i\omega = n \in \mathbb{Z}^+ $, where $n$ is a positive integer, the second solution becomes regular and must be kept and we end up with two independent solutions. One may argue that, even in this case, there is only one regular solution because the power of the two solutions differ by an integer, which leads to the appearance of a term of the form $\log (1-z)$ that should be discarded by the regularity condition. However, it turns out that  this $\log (1-z)$ term  vanishes for specific values of $k$, which we denoted as $k=k_{\{{n}\}}$. As a result, whenever the frequency and wave number are given by $(\omega, k) = (-i n, k_{\{n\}} )$ we have indeed two independent incoming solutions
\begin{equation} \label{lmnbg7}
    \phi(z) = \phi_0 + \phi_1(1-z) + \cdots +  (1-z)^n (\phi_n + \phi_{n+1}(1-z) + \cdots) \,,
\end{equation}
where $\phi_i (1 \le i \le n-1)$ and $\phi_i (n+1 \le i )$ are determined by $\phi_0$ and $\phi_n$, respectively. 
This implies that the Green's function is not uniquely defined at $(\omega, k) = (-i n, k_{\{n\}} )$ and has the form of $0/0$. These points are called `pole-skipping points' because the `would-be pole' is skipped by the zero in the numerator~\cite{Blake_2018,Blake:2018leo}.

To find out concrete conditions for the pole-skipping points, let us plug \eqref{lmnbg7} into \eqref{KGeq2} and solve it order by order.
We can express these equations collectively in a matrix form 
\begin{equation}
\label{scalleading}
\begin{split}
\setcounter{MaxMatrixCols}{20}
\mathbb{M}\cdot \Phi \equiv
&\left.\begin{pmatrix}
M_{00} && M_{01} && 0 && 0 && 0 &&  \cdots \\
M_{10} && M_{11} && M_{12} && 0  && 0 && \cdots \\
M_{20} && M_{21} && M_{22} && M_{23} && 0 && \cdots \\
\vdots && \vdots && \vdots && \vdots && \vdots && \ddots 
\end{pmatrix}
\begin{pmatrix}
\phi_0 \\ \phi_1 \\ \phi_2 \\ \phi_3 \\ \vdots
\end{pmatrix} = \begin{pmatrix}
0 \\ 0 \\ 0\\ 0 \\ \vdots
\end{pmatrix} \right. \,,\\
\end{split}
\end{equation}
where the $j$-th row is the equation of order $\mathcal{O}((1-z)^{j-1})$. For example, the leading order of equation is  
\begin{equation} \label{dksgs}
    M_{00} \phi_0 + M_{01} \phi_1 = 0 \,.
\end{equation}
The equations up to order $\mathcal{O}(1-z)$ is
\begin{align}
  &M_{00} \phi_0 + M_{01} \phi_1 = 0 \,, \\
  &M_{10} \phi_0 + M_{11} \phi_1 + M_{12} \phi_2 = 0 \,, 
\end{align}
which is equivalent to 
\begin{equation} \label{skdje}
       \frac{1}{M_{01}}(M_{01}M_{10} - M_{00}M_{11})\phi_0 + M_{12} \phi_2=0 \,
\end{equation}
by eliminating $\phi_1$. In general, the equation up to  order $\mathcal{O}((1-z)^n)$ for $(n \ge 1)$  is
\begin{equation}
    \label{pspeom1p}
    \frac{1}{M_{01}M_{12}\cdots  M_{(n-2)(n-1)}}\det \mathcal{M}^{(n)}(\omega,k)\,\phi_0 +  M_{(n-1)n}\,\phi_n=0 \,,
\end{equation}
where $\mathcal{M}^{(n)}$ is the $n \times n$ sub-matrix in the top-left corner of the matrix $\mathbb{M}$.

In our case, the matrix element $M_{ij}$ has the form
\begin{align}
    M_{ij} = a_{ij} i\omega + b_{ij} k^2 + c_{ij} \,,
\end{align}
where $a_{ij}, b_{ij},$ and  $c_{ij}$ are real numbers. In particular,
\begin{equation} \label{inytg}
   M_{(j-1)j} = j-i\omega \,,\quad (j \ge 1)  \,,
\end{equation}
and
\begin{equation}
\label{scalelem}
\begin{split}
&M_{00} = -\frac{1}{2}\left((d-1) i\omega +{k}^2+\left(\frac{d-2}{2}\right)^2+\Delta(\Delta-d)\right)\,,\\
&M_{10} = \frac{1}{2}\left(\frac{(d-1)}{2} i\omega + {k}^2+\left(\frac{d-2}{2}\right)^2\right)\,,\\
&M_{11} = -\frac{1}{2}\left(\frac{(d-5)}{2} i\omega+{k}^2+\left(\frac{d-2}{2}\right)^2+\frac{\Delta(\Delta -d)}{2} -(d-4)\right)\,,
\end{split}
\end{equation}
where we used the relation $m^2 = \Delta(\Delta-d)$. 

\begin{figure}
\centering
    \subfigure[$\Delta = 4.2\,(\delta=2.2)$]{\includegraphics[width=4.8cm]{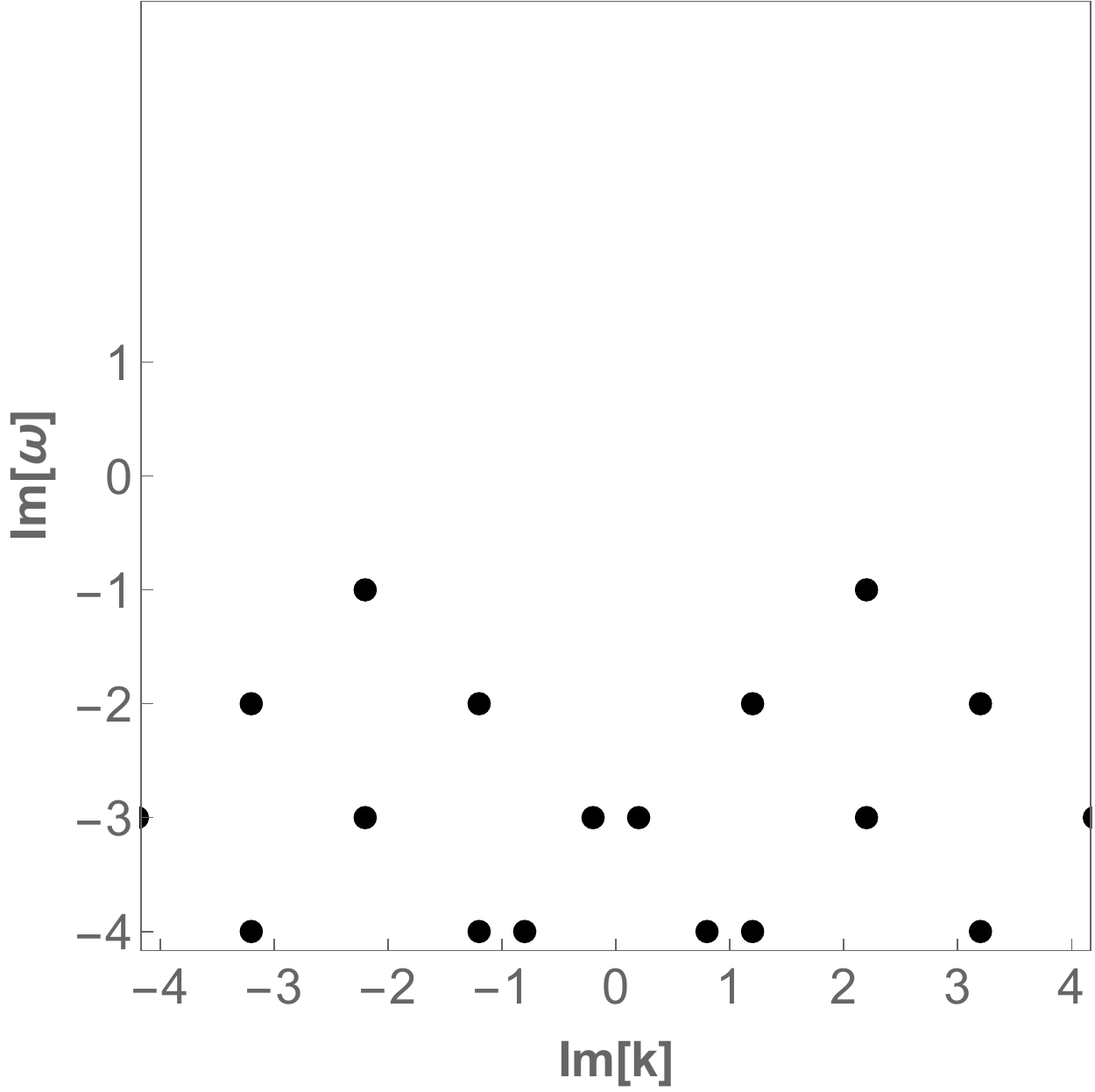}
    \label{holscalpsploca}
    }
    \subfigure[$\Delta = 4\,(\delta=2.0)$]{\includegraphics[width=4.8cm]{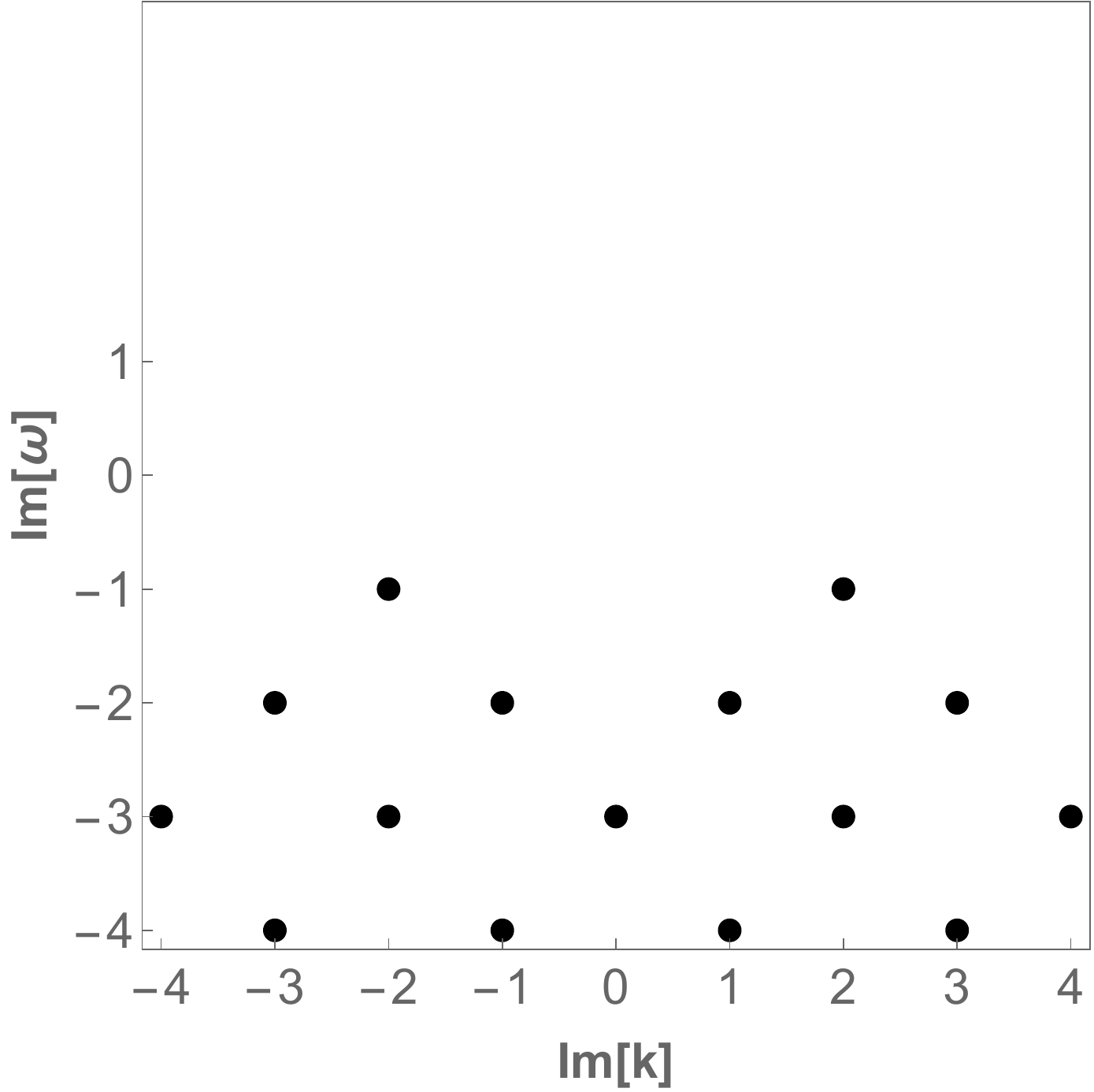}
    \label{holscalpsplocb}
    }
    \subfigure[$\Delta = 3.8\,(\delta=1.8)$]{\includegraphics[width=4.8cm]{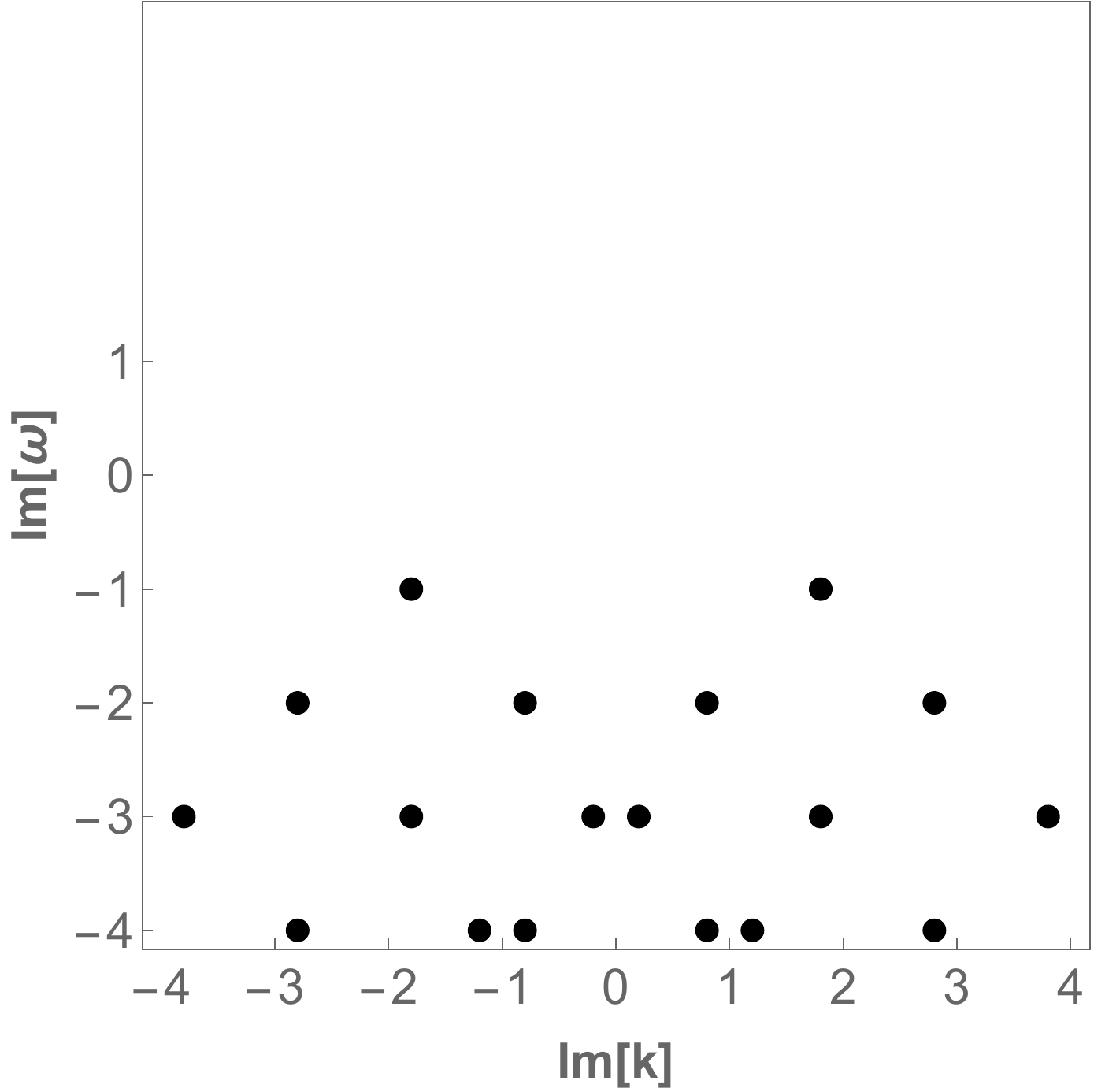}
    \label{holscalpsplocc}
    }
\caption{The black dots represent the pole-skipping points at $d=4$ (obtained from \eqref{scalleading}) of a scalar field propagating in a Rindler-AdS geometry. Here $\delta:=\Delta-d/2$.}
\label{holscalpsploc}
\end{figure}

Using \eqref{inytg}, the general equation \eqref{pspeom1p} becomes 
\begin{equation}
    \label{pspeom1}
    \frac{1}{\mathcal{N}^{(n)}(\omega)}\det \mathcal{M}^{(n)}(\omega,k)\,\phi_0 +  (n-i\omega)\,\phi_n=0 \,, \quad (n \ge 1) \,,
\end{equation}
where\footnote{This is closely related with $M_{n-1, n-1} = 2 n T - 2 i \omega$, $T =\frac{1}{2\pi}$. See Equation \eqref{scalelem} and Equation  \eqref{HawkingT}.}
\begin{equation}
 \mathcal{N}^{(n)}(\omega) \equiv \prod_{m=1}^{n-1}(i\omega - m ) \,.
\end{equation}
The pole-skipping points are obtained when $\phi_0$ and $\phi_n$ are independent, which is realized if the point  $(\omega, k) = (\omega_n, k_{\{n\}})$ is such that 
\begin{equation}
    \omega_n = -i n \,, \qquad \det \mathcal{M}^{(n)}(\omega_n,k_{\{n\}}) = 0 \,.
\end{equation}
For a given $\omega_n$, the determinant $\det \mathcal{M}^{(n)}(\omega_n,k)$ is a polynomial in $k$ of order $2n$. This implies that there are $2n$ solutions for $k$. We collectively denote them by $k_{\{n\}}$. 
For example,
\begin{equation} \label{sleadpsp}
\begin{split}
&\omega_{1}=-i\,, \quad k_{\{1\}}=\pm i\left(\Delta-\frac{d}{2} \right)\,,\\
&\omega_{2}=-2i\,, \quad k_{\{2\}}=\pm i\left(\Delta-\frac{d}{2}\pm1\right)\,.
\end{split}
\end{equation}
Figure \ref{holscalpsploc} shows the `tower' of pole-skipping points of a scalar field propagating in \eqref{AdSRind}.

\begin{figure}
\centering
    \subfigure[ $\Delta=4.2\, (\delta=2.2)$ ]{\includegraphics[width=4.9cm]{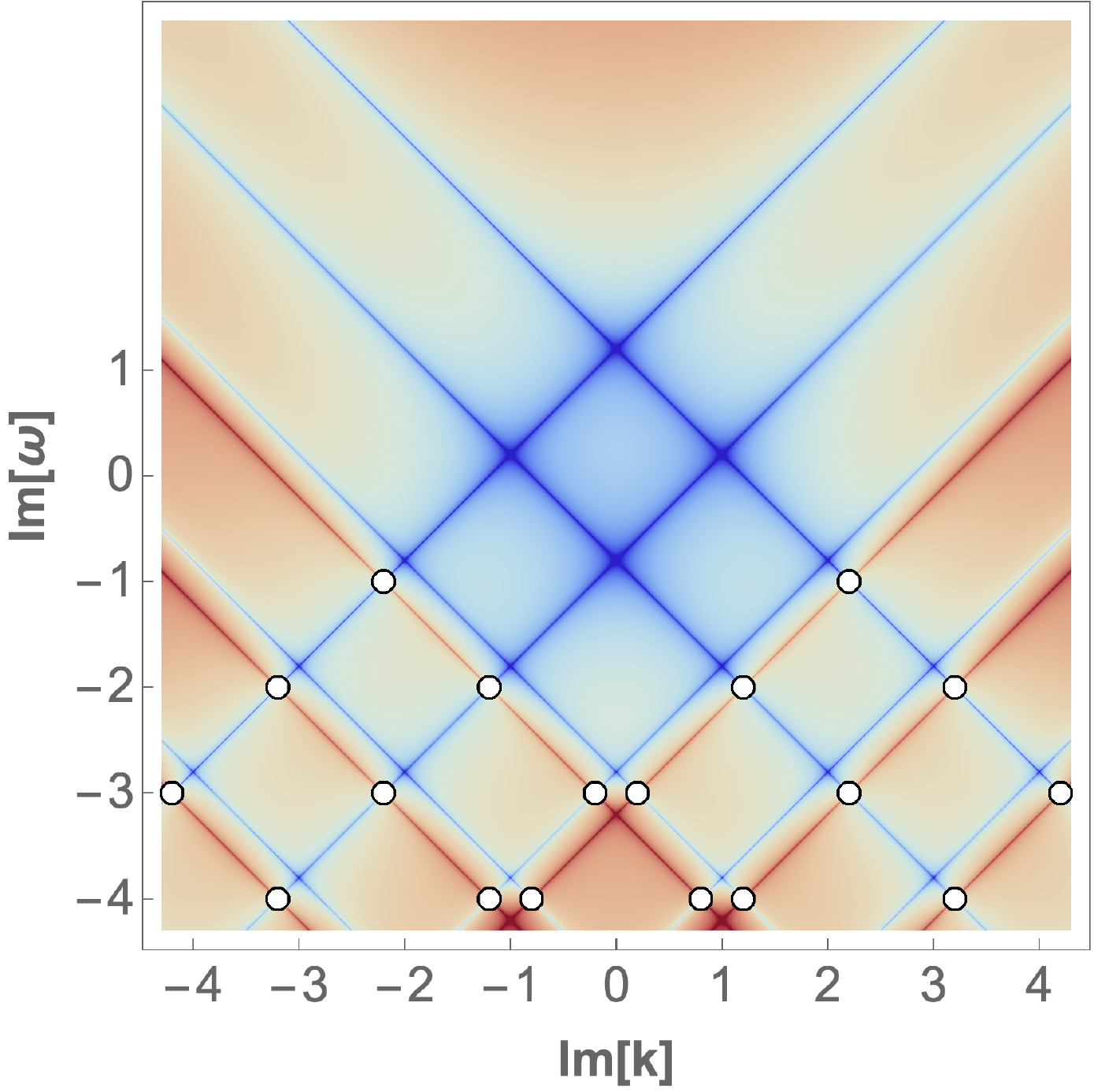}\label{fig:regscalar4.2}}
    \subfigure[ $\Delta=4.0\, (\delta=2.0)$]{\includegraphics[width=4.9cm]{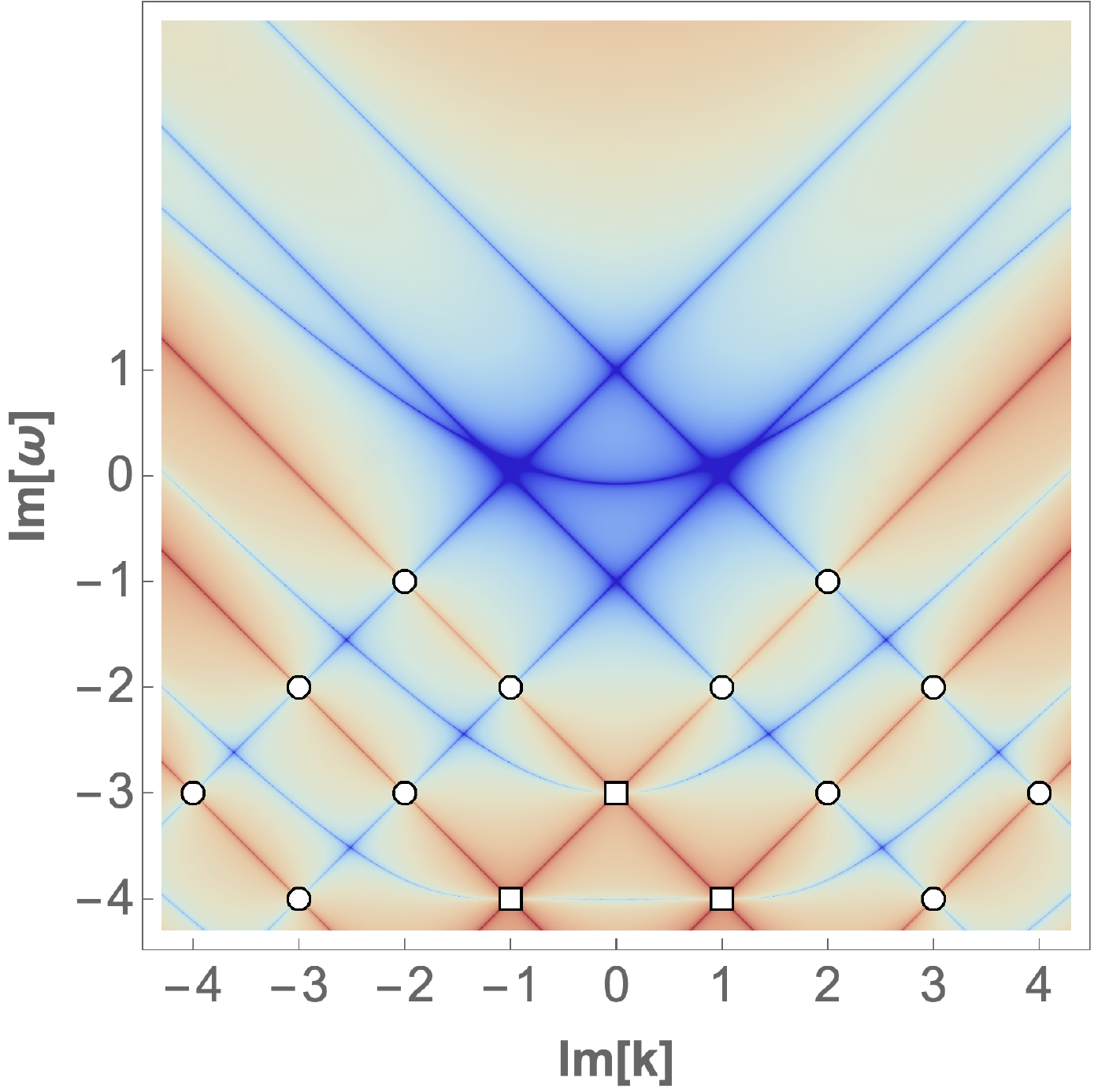}\label{fig:regscalar4}}
    \subfigure[$\Delta=3.8\, ( \delta=1.8)$]{\includegraphics[width=4.9cm]{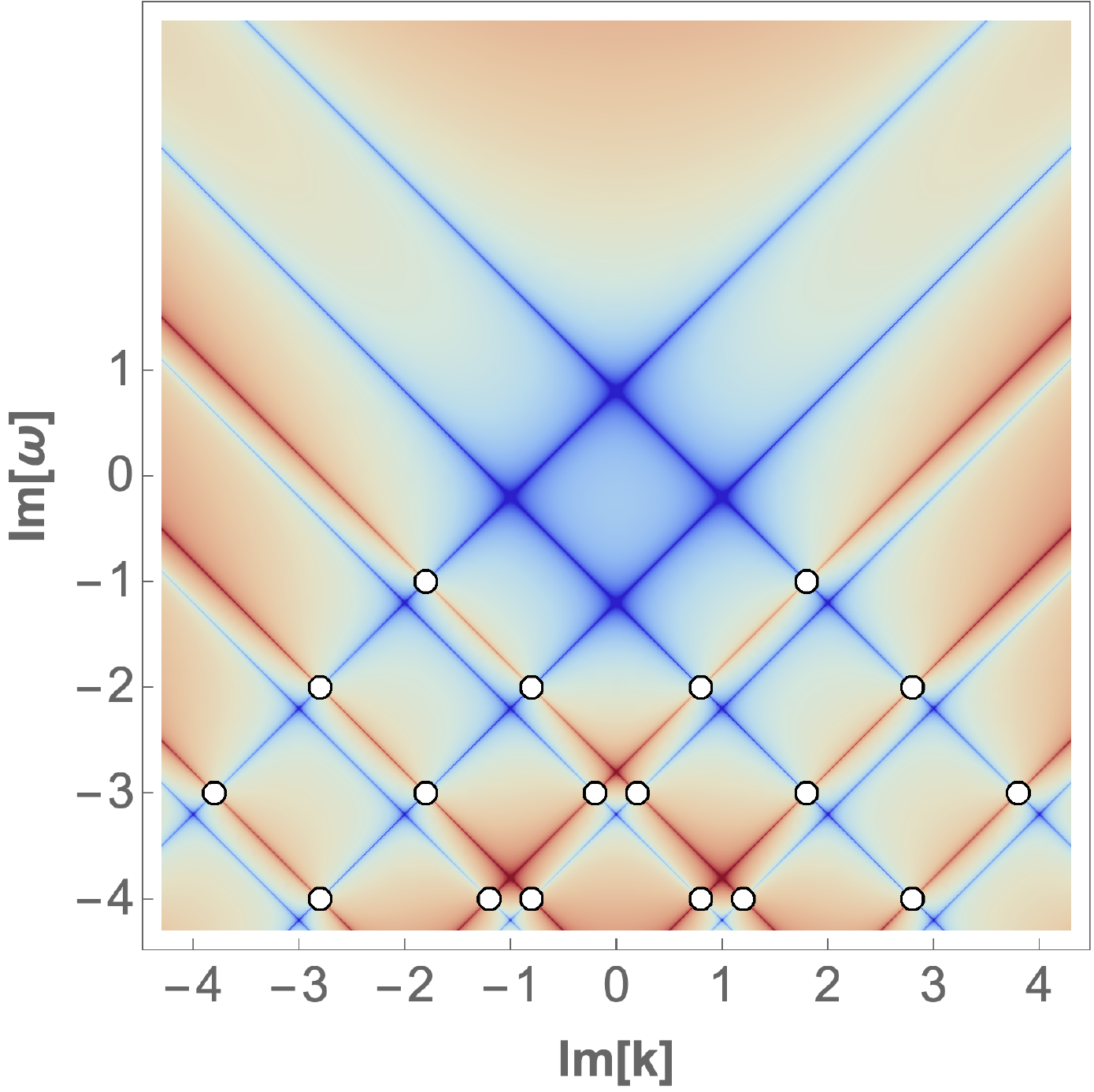}\label{fig:regscalar3.8}}
    \subfigure[ $\Delta=4.5\,(\delta=2.5)$ ]{\includegraphics[width=4.9cm]{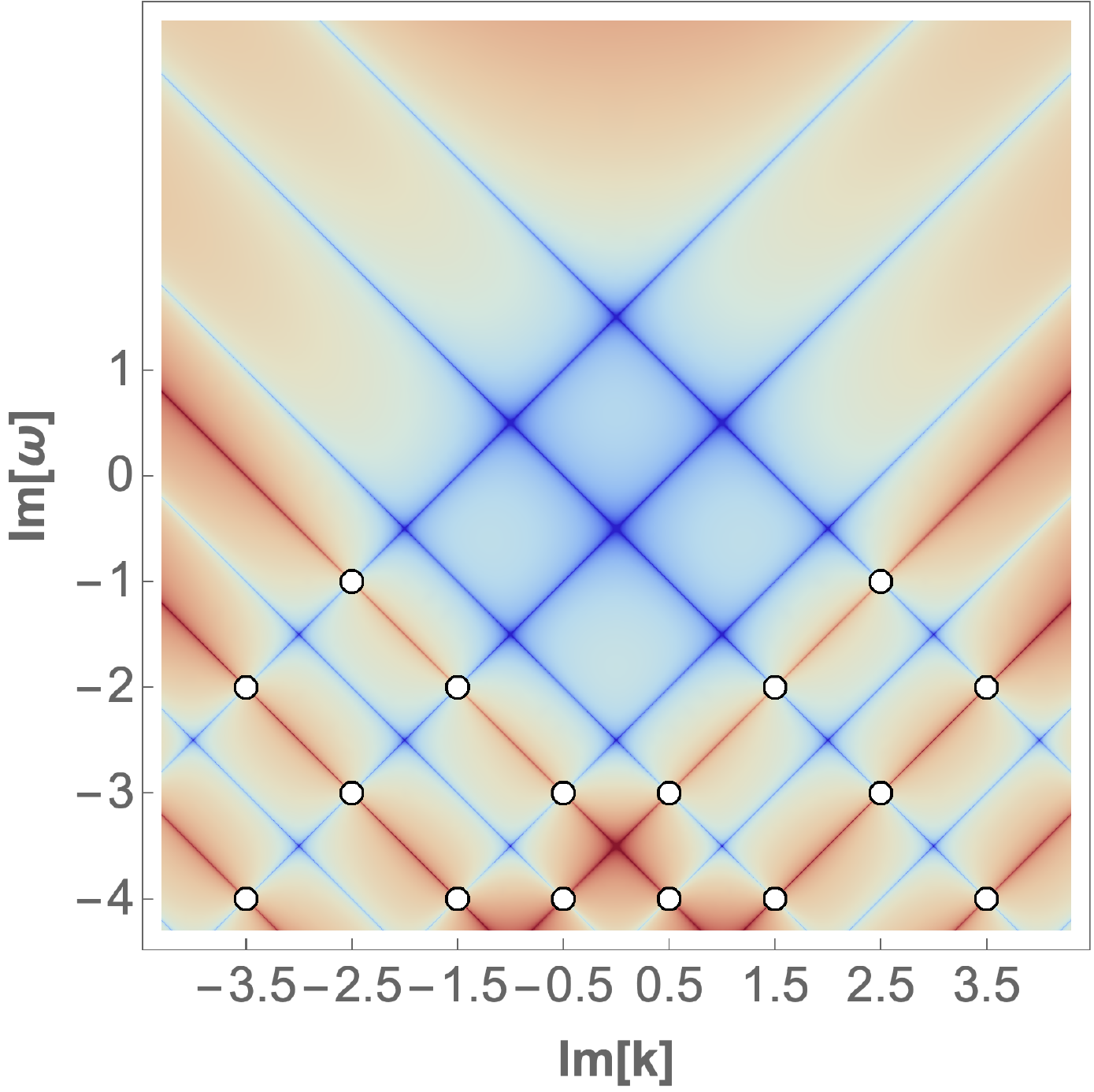}\label{fig:regscalar4.5}}
    \subfigure[ $\Delta=2\, (\delta=0)$ ]{\includegraphics[width=4.9cm]{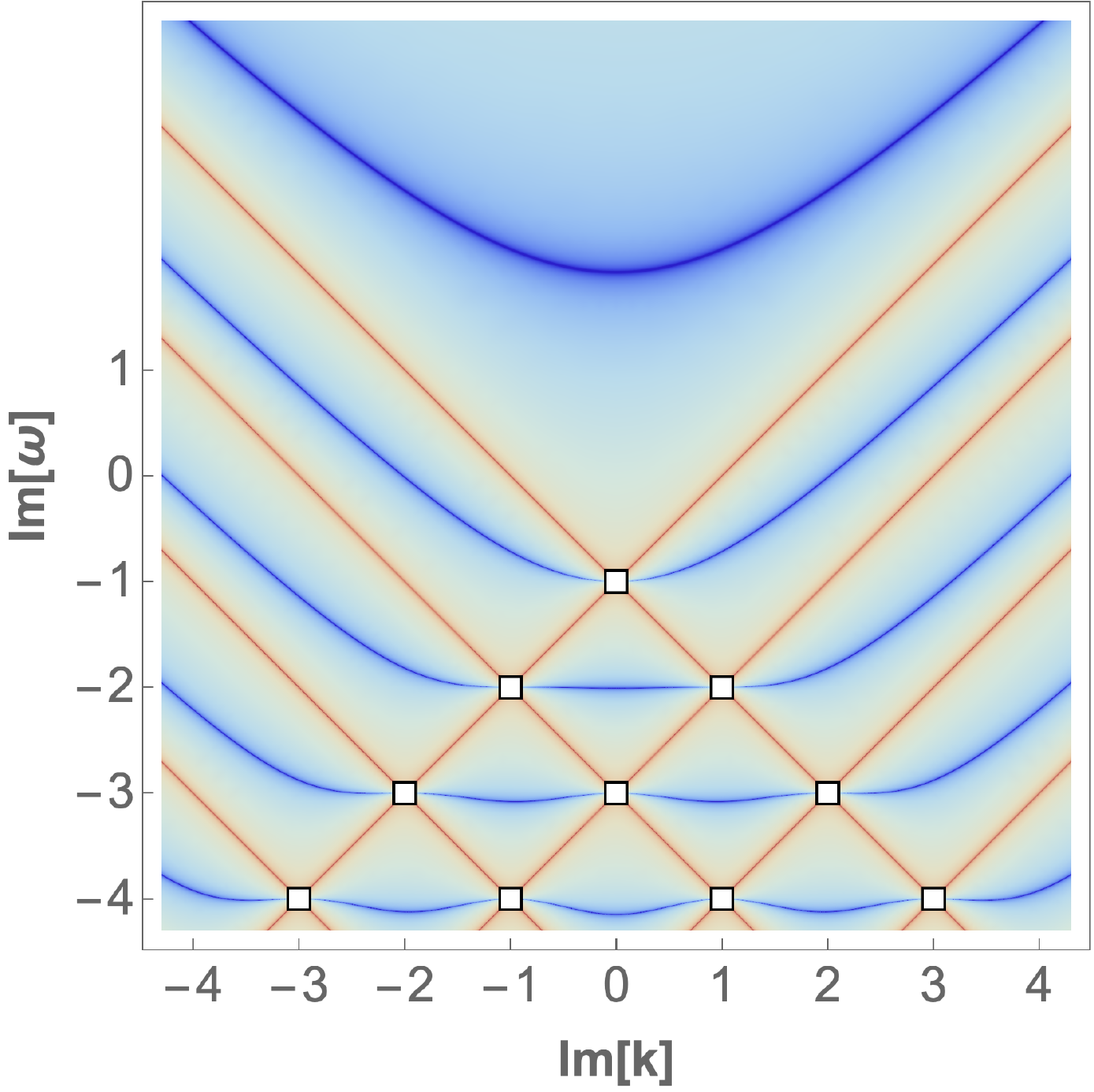}\label{fig:regscalar2.0}}
    \caption{$\log |\mathcal{G}^\Delta(\mathrm{Im}[\omega],\mathrm{Im}[k])|$ at $d=4$. 
    The blue lines and red lines represent zeros and poles of the Green's function, respectively. The white circles and squares are type I and type II pole-skipping points, respectively. The meaning of the type I and II classifications will be explained in section \ref{section3}.
    } \label{fig:regscalar}
\end{figure}


\subsubsection{Field theory results}

In this section, we briefly summarize the result for the retarded Green's function of scalar operators and the corresponding pole-skipping points computed for a  CFT living in $S^1\times\mathbb{H}^{d-1}$~\cite{Ahn:2020bks}.

The analytic form of the Green's function of a scalar field with conformal dimension $\Delta$ in momentum space is conveniently expressed in terms of $\delta:=\Delta-d/2 >-1$\footnote{This restriction comes from the unitarity bound on the conformal dimension of the scalar operator $\delta\geq-1$ and the non-existence of pole-skipping point at $\delta=-1$, which was revealed in \cite{Ahn:2020bks}.}. We refer to \cite{Ahn:2020bks} for a detailed derivation and here we show only the final results. 
For non-integer $\delta$
\begin{equation}
\mathcal{G}^\Delta(\omega,k)\propto\frac{\Gamma(x +\delta/2)}{\Gamma(x-\delta/2)}\frac{\Gamma(y +\delta/2)}{\Gamma(y-\delta/2)}\Gamma(-\delta)\,,\label{fourierscalar2pt}
\end{equation}
and for non-negative integer $\delta$ $\in \{\mathbb Z^+, 0\}$
\begin{equation}
    \mathcal{G}^\Delta(\omega,k)\propto\frac{\Gamma(x +\delta/2)}{\Gamma(x-\delta/2)}\frac{\Gamma(y +\delta/2)}{\Gamma(y-\delta/2)}\left[\psi(x+\delta/2)+\psi(y+\delta/2)\right]\,,\label{eq-Gscalar-1}
\end{equation}
where 
\begin{equation}
    x= \frac{-i\omega + ik + 1}{2}\,, \quad y= \frac{-i\omega - ik + 1}{2} \,,\label{xydef}
\end{equation}
and $\psi$ is the digamma function.

Note that even though these are derived in \cite{Ahn:2020bks}, the case with non-negative integer $\delta$ $\in \{\mathbb Z^+, 0\}$  \eqref{eq-Gscalar-1} was not analyzed there and important subtleties due to the digamma functions were not revealed. Here, we will make this point clear.

We may analytically identify the zeros and poles of the Green's function by investigating the properties of the gamma functions in \eqref{fourierscalar2pt} and \eqref{eq-Gscalar-1} in the two-dimensional space of Im$[\omega]$ versus Im$[k]$. For an easy visualization and cross check, we make numerical plots of \eqref{fourierscalar2pt} and \eqref{eq-Gscalar-1}. See, for example, Fig. \ref{fig:regscalar}, where we consider the case $d=4$ and $\Delta=2,\ 3.8,\ 4,\ 4.2,\ 4.5$. For clear visualization, we make plots of the logarithm of the absolute values of Green's function, $\log |\mathcal{G}^\Delta(\mathrm{Im}[\omega],\mathrm{Im}[k])|$. 
The red lines represent pole conditions and the blue lines (curves) represent zero conditions. The points where the red and blue lines cross each other are pole-skipping points. They are represented by white circles (type-I) or white squares (type-II). We will explain the difference between these two types of pole-skipping point in section \ref{section3}.

To understand the origin of the zeros and poles from the analytic expressions \eqref{fourierscalar2pt} and \eqref{eq-Gscalar-1},
let us consider three cases depending on the values of $\delta$ separately.  
\paragraph{Case 1: non-integer $\delta$} There is no cancellation between the gamma functions in  \eqref{fourierscalar2pt}. Thus, from the gamma functions in the numerator and denominator, the non-positive integer value of the argument $x\pm\delta/2$ (or $y\pm\delta/2$) give an infinite number of zeros and poles. For example, if $\delta=2.2, 1.8$ or $\delta=2.5$, an infinite number of straight blue and red lines spread out beyond the range of Fig. \ref{fig:regscalar4.2}, \ref{fig:regscalar3.8}, and \ref{fig:regscalar4.5}.

\paragraph{Case 2: $\delta\in\mathbb Z^+$} The first factor of the gamma functions in \eqref{eq-Gscalar-1}  becomes a $\delta$-th order polynomials of $x$ and $y$:
\begin{equation} \label{con100}
    {\underbrace{(x+\delta/2-1)(x+\delta/2-2)\cdots(x-\delta/2)}_{\delta\ \mathrm{factors}} \cdot {(x \to y)} \,,}
\end{equation}
which gives a conditions for zeros.  For example, if $\delta=2$ this condition gives four (two from `$x$', two from `$y$') straight blue lines with slopes $\pm 1$ in Fig. \ref{fig:regscalar4}. 
The next two digamma functions in \eqref{eq-Gscalar-1} give the conditions for poles:
\begin{equation} \label{con200}
    {-i\omega \pm ik+\delta+1=-2j,\qquad(j=0,1,2\dots)\,.}
\end{equation}
 For example, if $\delta=2$ this condition gives the straight red lines in Fig. \ref{fig:regscalar4}.
Importantly, there is another type of zero-conditions, which comes from the cancellation between two digamma functions, i.e. 
\begin{equation} \label{con300}
 \psi(x+\delta/2) + \psi(y + \delta/2) = 0   \,. 
\end{equation}
These conditions can not be written analytically but can be shown graphically by plotting $\psi(x + \delta/2) + \psi(y + \delta/2)$. They are represented by the blue curves in Fig. \ref{fig:regscalar4}.
\paragraph{Case 3: $\delta = 0$} 
The Green's function \eqref{eq-Gscalar-1} becomes the summation of two digamma functions. 
\begin{equation} \label{con400}
 \psi(x) + \psi(y) = 0   \,, 
\end{equation}
from which all zero and pole conditions appear. 
Fig. \ref{fig:regscalar2.0} shows this case. Note that there is no straight blue lines. All blue curves come from the summation of two digamma functions. The red straight lines appear if $x$ and $y$ are non-positive integers.

All things considered, the pole-skipping points of the scalar Green's function can be summarized as in \cite{Ahn:2020bks}:
\begin{equation}  
    \omega_{n}=-in, \ \ \mathrm{and}\ \ 
    k_{\{n\}}=\pm i\left(-n+2q+\delta-1\right)\label{sps}\,,
\end{equation}
where $n=1,2,\cdots$, $q=1,2,\cdots,n$, and  $\delta>-1$. 
Here, the subscript $\{n\}$ of $k$ includes the $q$ dependence collectively. For example, for the first two cases $n=1,2$  \eqref{sps} read
\begin{equation}
\begin{split}
&\omega_{1}=-i\,, \quad k_{\{1\}}=\pm i\left(\Delta-\frac{d}{2} \right)\,,\\
&\omega_{2}=-2i\,, \quad k_{\{2\}}=\pm i\left(\Delta-\frac{d}{2}\pm1\right)\,,
\end{split}
\end{equation}
which agree with the results from the near horizon analysis \eqref{sleadpsp}. The agreements can be also confirmed by comparing Fig. \ref{holscalpsploc} (for the near horizon analysis) and Figs. \ref{fig:regscalar4.2}, \ref{fig:regscalar4}, \ref{fig:regscalar3.8} (for the field theory Green's function). The pole-skipping points in both cases exactly match.

Before closing this section we want to recall the dual role of the two digamma functions in \eqref{eq-Gscalar-1}, which give conditions for zeros as well as for poles.
In~\cite{Blake:2019otz}, the pole-skipping points of a scalar field in the BTZ background were computed. The mathematical structure of the Green's function there is similar to our case. The authors of ~\cite{Blake:2019otz} claim that some of the pole-skipping points are `anomalous' partly\footnote{Another reason to call it `anomalous' is that the Green's function near the pole-skipping point does not depend on $\delta \omega/\delta k$ along a linear path. This will be explained in detail in Sec.\ref{section3}.} because they arise as an intersection of two lines of poles without any line of zeros. See Fig.~4 in~\cite{Blake:2019otz}. 
There is, however, a line of zeros that passes through these points. In doing the plots of the pole-skipping points, one needs to consider the zero-conditions that comes from the cancellation between the two digamma functions, as in \eqref{con300}. By taking that into consideration, Fig.~4 in ~\cite{Blake:2019otz} becomes the same as   Fig.~\ref{fig:regscalar4} because \eqref{eq-Gscalar-1} with $\Delta=4, d=4$ is the same as (C.5) in \cite{Blake:2019otz} with $\Delta=3$.  This may mean that this point is not very `anomalous' since it is also defined as the point where a line of zeros intersects with lines of poles. We call this type of point a type-II pole-skipping point, and we discuss its properties in more detail in Sec.\ref{section3}.

\subsection{Vector field} 
\subsubsection{Near horizon analysis}

In this section, we obtain  the pole-skipping points of a massless vector field in the hyperbolic black hole geometry \eqref{inEFhypBH}. We assume the following action
\begin{align}
\begin{split}
&S_{A} = -\frac{1}{4}\int \dd^{d+1}x \sqrt{-g} \, F^2\,,\label{paction}
\end{split}
\end{align}
where $F_{\mu\nu} := \partial_{\mu} A_{\nu} - \partial_{\nu} A_{\mu}$. The corresponding equation of motion is
\begin{equation}
\label{eqMaxwell}
\nabla_{\mu}F^{\mu \nu} = 0\,.
\end{equation}

In hyperbolic space, a general perturbation of the bulk vector field $A_\mu$ can be decomposed into ``longitudinal''  and ``transverse'' channels. More explicit expressions of each channels can be seen after decomposing the metric \eqref{inEFhypBH}
\begin{equation}
\label{incomingEF}
\begin{split}
\dd s^2&=-\frac{f(z)}{z^2}\dd v^2-\frac{2}{z^2} \dd v \dd z+\frac{1}{z^2}\left( \dd\chi^2+\sinh^2\chi \dd\Omega_{d-2}^2 \right)\,, \\
f(z)&=1-z^2
\,.
\end{split}
\end{equation}
into the following form:
\begin{align}
\label{metricdecom}
\dd s^2 = g_{ab}\dd y^a \dd y^b + \frac{1}{z^2} \gamma_{ij}\dd x^i \dd x^j,
\end{align}
where $y^1 = v$, $y^2=z$, and $\gamma_{ij} \dd x^i \dd x^j =\dd \chi^2+\sinh^2\chi \,\dd \Omega_{d-2}^2 = \dd H_{d-1}^2$.
The vector field in terms of ``longitudinal channel'' $(A^L_a,A^L)$ and ``transverse channel'' $A_i^T$ can be written as~\cite{Ueda:2018xvl}:
\begin{equation}
\label{decomvec}
A_{\mu}dx^{\mu} = A^L_a dy^a + \hat{D}_iA^Ldx^i + A_i^T dx^i, \quad \hat{D}^iA_i^T = 0\,,
\end{equation}
where the differential operator $\hat{D}_i$ denotes the covariant derivative with respect to  $\gamma_{ij}$.

The equations of motion for the two channels are decoupled so one can consider each channel separately. Here we focus on the ``longitudinal channel'' and set $A^T_i=0$ in \eqref{decomvec}. The ``transverse channel'' is similar to the scalar field case \ref{scalar}.
 
 

For simplicity, we consider perturbations that do not depend on the coordinates $\theta_i$ on $S^{d-2}$, i.e., $A_{\mu} = A_{\mu} (v,z,\chi)$. 
Let us consider the components $A_\mu$ in momentum space:
 \begin{equation}
 \label{diffansatz}
 \begin{split}
    &A_v = A_v^L(v,z,\chi)= \int\dd \omega \,\dd k \,\tilde{A}_v(z) \,e^{-i \omega v}\mathbb{S}_{k}(\chi)\,,\\
    &A_z = A_z^L(v,z,\chi) = \int\dd \omega\, \dd k\, \tilde{A}_z(z) \, e^{-i \omega v}\mathbb{S}_{k}(\chi)\,,\\
    &A_\chi = \hat{D}_{\chi} A^L(v,z,\chi) = \int\dd \omega\, \dd k\, \tilde{A}_\chi(z )\, e^{-i \omega v}\mathbb{S}_{k}'(\chi)\,,
 \end{split}
 \end{equation}
where $\mathbb{S}_{k}$ is the scalar harmonics defined in \eqref{EVeq1}:
\begin{equation}
\label{EVeq}
\left(\square_{H_{d-1}}+{k}^2+\left(\frac{d-2}{2}\right)^2\right)\mathbb{S}_{k}(\chi)=0.
\end{equation}

The Maxwell equations \eqref{eqMaxwell} yield three   coupled equations for $\tilde{A}_v(z), \tilde{A}_z(z)$, and $\tilde{A}_\chi(z)$. They can be expressed in terms of two  equations for the gauge invariant quantities
\begin{equation}
    \mathcal{U}_v(z)\equiv\tilde{A}_v(z)+i\omega\tilde{A}_\chi(z)\,, \qquad \mathcal{U}_z(z)\equiv \tilde{A}_z(z) - \tilde{A}'_\chi(z) \,.
\end{equation}
For $\omega \ne 0$ and $k^2+\left(\frac{d-2}{2}\right)^2 \ne 0$, the Maxwell equations reduce to
\begin{equation}
\label{diffeq}
\begin{split}
&\mathcal{U}_z=\frac{i \omega  \mathcal{U}_v'(z)-\left(k^2 + \left(\frac{d-2}{2}\right)^2\right) \mathcal{U}_v(z)}{\omega ^2-\left(k^2 + \left(\frac{d-2}{2}\right)^2\right) f(z)}\,,\\
   &\mathcal{U}_v''(z)+\Omega(\omega,k)\mathcal{U}_v'(z)+\Xi(\omega,k)\mathcal{U}_v(z)=0\,,\\
   &\Omega(\omega,k) := \frac{\omega^2 f'(z)}{\left(\omega^2 - \left(k^2+\left(\frac{d-2}{2}\right)^2\right)f(z)\right)f(z)}+\frac{2i \omega}{f(z)}-\frac{d-3}{z}\,,\\
   &\Xi(\omega,k):=\frac{i \omega\left(k^2+\left(\frac{d-2}{2}\right)^2\right)f'(z)}{\left(\omega^2 - \left(k^2+\left(\frac{d-2}{2}\right)^2\right)f(z)\right)f(z)}-\frac{\left(k^2+\left(\frac{d-2}{2}\right)^2\right)z+i \omega(d-3)}{z f(z)}\,,
\end{split}
\end{equation}
For $\omega=0$ and $k^2+\left(\frac{d-2}{2}\right)^2 = 0$, they become
\begin{equation}
\label{eq:vectorzeroomega}
    \mathcal{U}_v''(z)-\frac{2}{z}\mathcal{U}_v'(z)=0 \,.
\end{equation}

\begin{figure}
 \centering
    \subfigure[$d=4, \Delta = 3$]{\includegraphics[width=4.8cm]{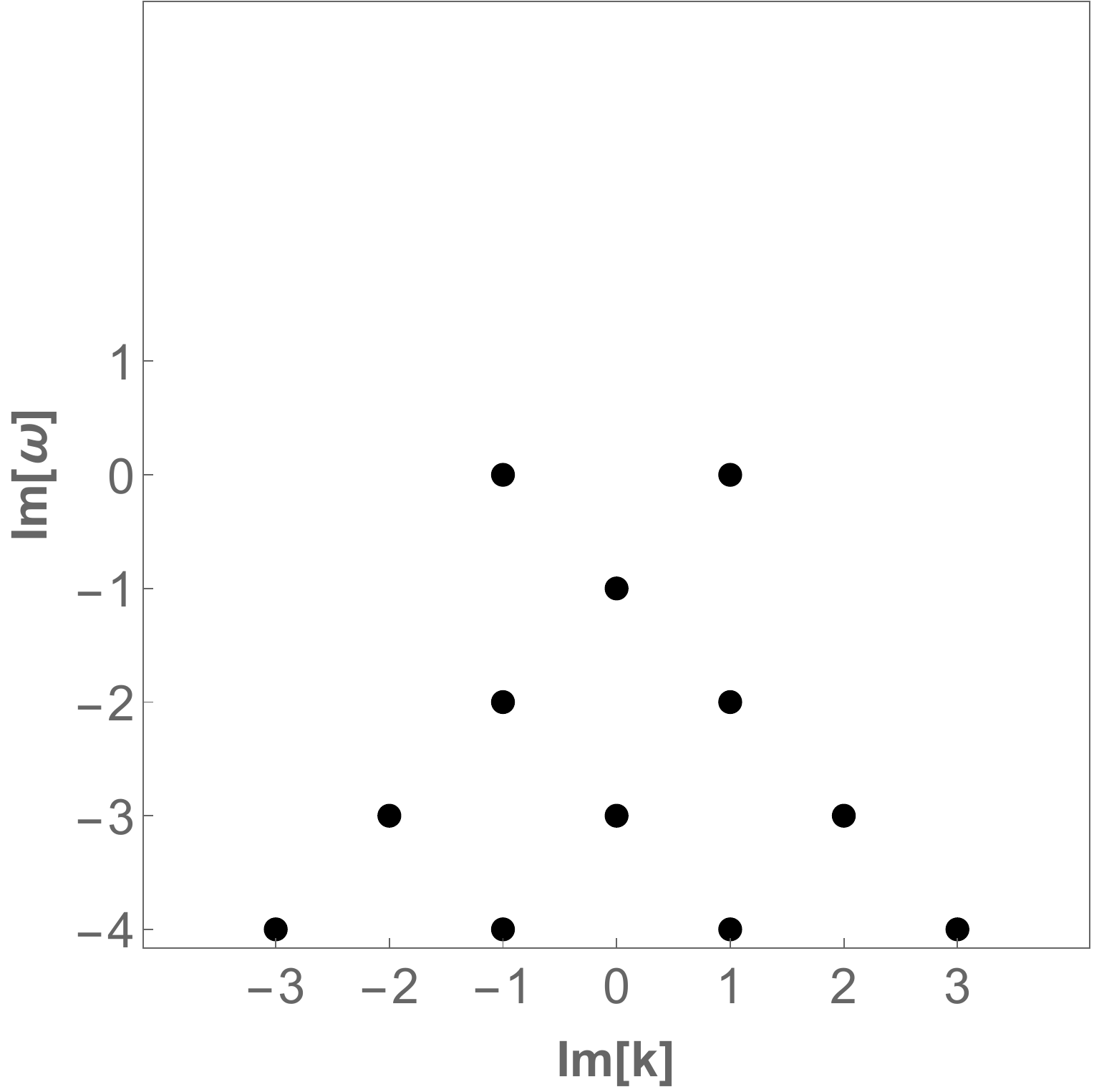}
    \label{diffpspsap}}
    \subfigure[$d=5, \Delta = 4$]{\includegraphics[width=4.8cm]{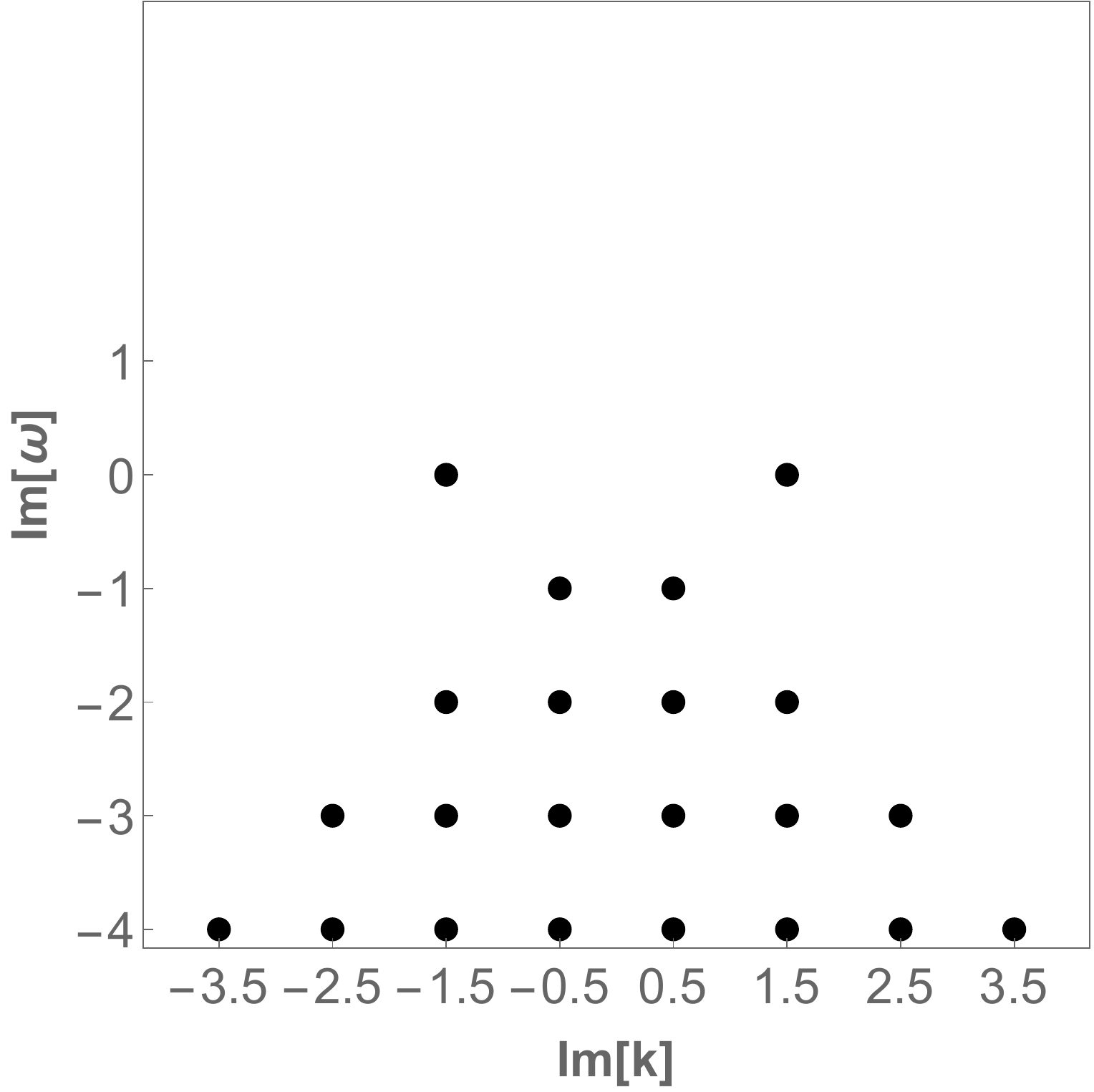}
    \label{diffpspsbp}}
    \subfigure[$d=6, \Delta = 5$]{\includegraphics[width=4.8cm]{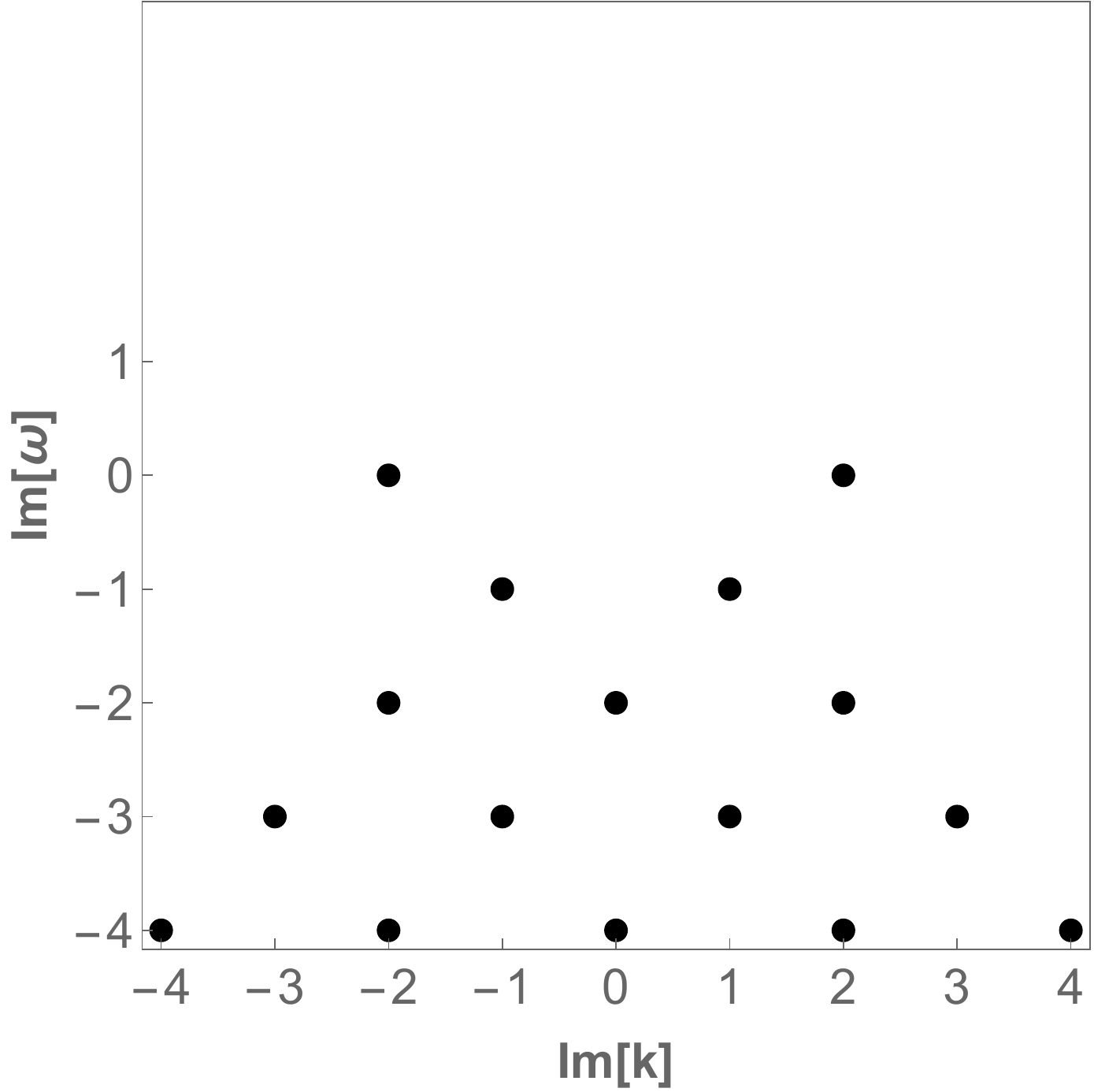}
    \label{diffpspscp}}
    \caption{Locations of the pole-skipping points of longitudinal channels.}
    \label{diffpsps}
\end{figure}

Let us start with the case in which $\omega=0$ and $k^2+\left(\frac{d-2}{2}\right)^2=0$.  Eq.~\eqref{eq:vectorzeroomega} implies that the near horizon behavior of $\mathcal{U}_v$ is
\begin{equation}
\label{eq:zeroomegaasymp}
\mathcal{U}_v(z) = \mathcal{U}_v^{(0)} + \mathcal{U}_v^{(1)}(1-z) + \ldots \,, 
\end{equation}
where $\mathcal{U}_v^{(0)}$ and $\mathcal{U}_v^{(1)}$ are independent of each other. Thus the points $\omega=0, k^2+\left(\frac{d-2}{2}\right)^2=0$ are pole-skipping points since \eqref{eq:zeroomegaasymp} has two independent regular solutions .
For $\omega \ne 0$, it turns out that the formalism to obtain pole-skipping points, which was developed for the scalar field, works also for the vector field. The result of the indicial equation is the same as \eqref{gscgs}  so we only need to identify the matrix element in \eqref{scalleading}. Instead of repeating the procedure, we show the final results for the pole-skipping points obtained using the near horizon analysis when $d=4, d=5$, and $d=6$ in Fig. \ref{diffpsps}. Some numerical values are summarized in the table below
\begin{equation}\label{VNHApsps}
\begin{array}{|c|c|c|}
\hline
 d = 4, \Delta = 3  & d = 5, \Delta = 4 & d = 6, \Delta = 5 \\
\hline
 \quad {\omega}_{1}=0\,, \quad {k}_{\{1\}}=\pm i \quad&\quad {\omega}_{1}=0\,, \quad {k}_{\{1\}}=\pm \frac{3i}{2} \quad &\quad {\omega}_{1}=0\,, \quad {k}_{\{1\}}=\pm 2i \quad \\
 {\omega}_{2}=-i\,, \quad {k}_{\{2\}}=0 & {\omega}_{2}=-i\,, \quad {k}_{\{2\}}=\pm \frac{i}{2} & {\omega}_{2}=-i\,, \quad {k}_{\{2\}}=\pm i \\
 \vdots & \vdots & \vdots \\
\hline
\end{array}
\end{equation}
where $\Delta = d-1$, since we are considering a massless vector field \footnote{The scaling dimension $\Delta$  for a massive vector field in the holographic dual satisfies the relation: $m^2 = \left(\Delta-1\right)\left(\Delta - d +1\right)$.}.
 \begin{figure}
 \centering
     \subfigure[$d=4, \Delta=3\,(\delta'=0)$]{\includegraphics[width=4.8cm]{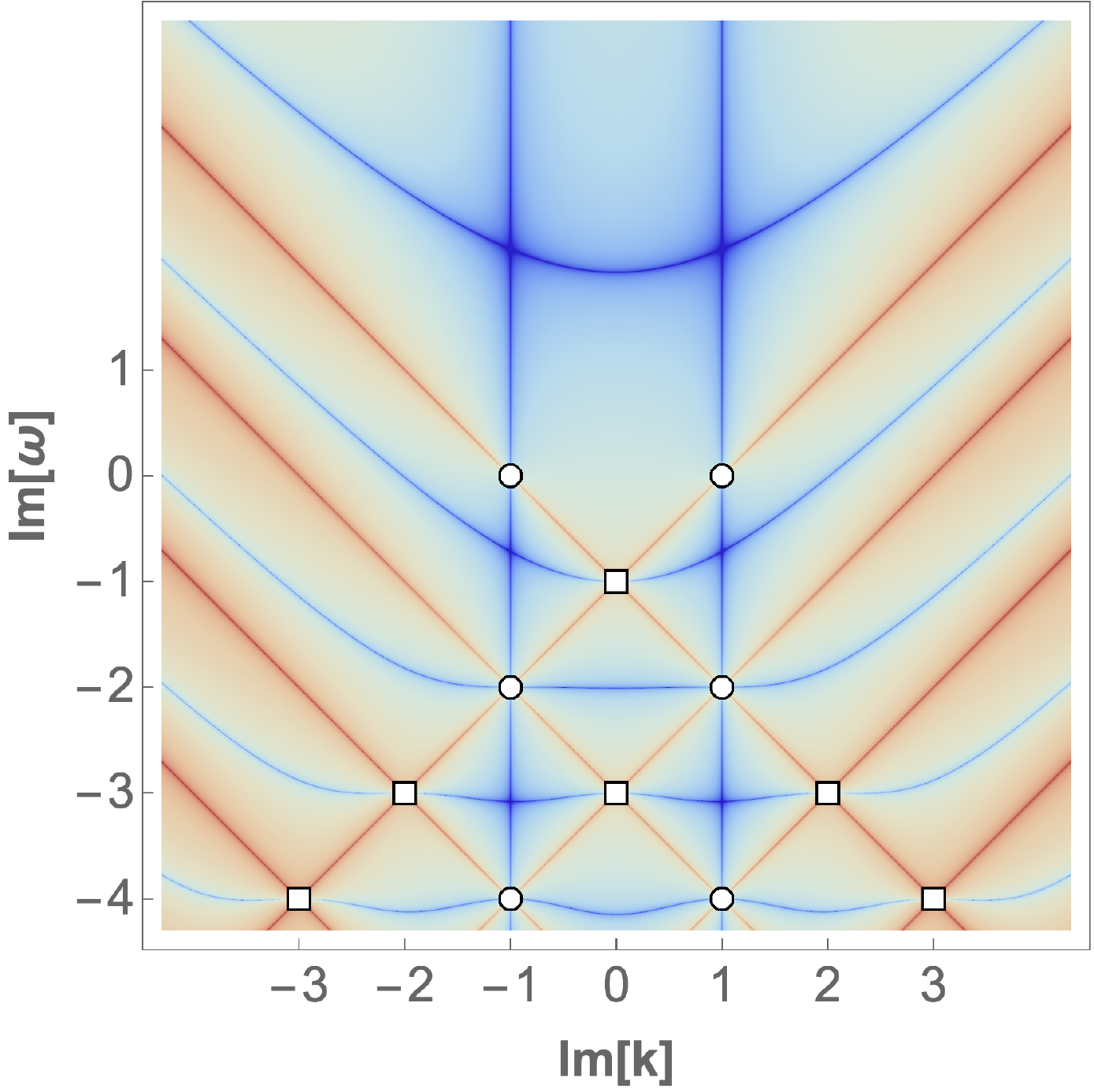}\label{massless123a}}
     \subfigure[$d=5, \Delta=4\,(\delta'=0.5)$]{\includegraphics[width=4.8cm]{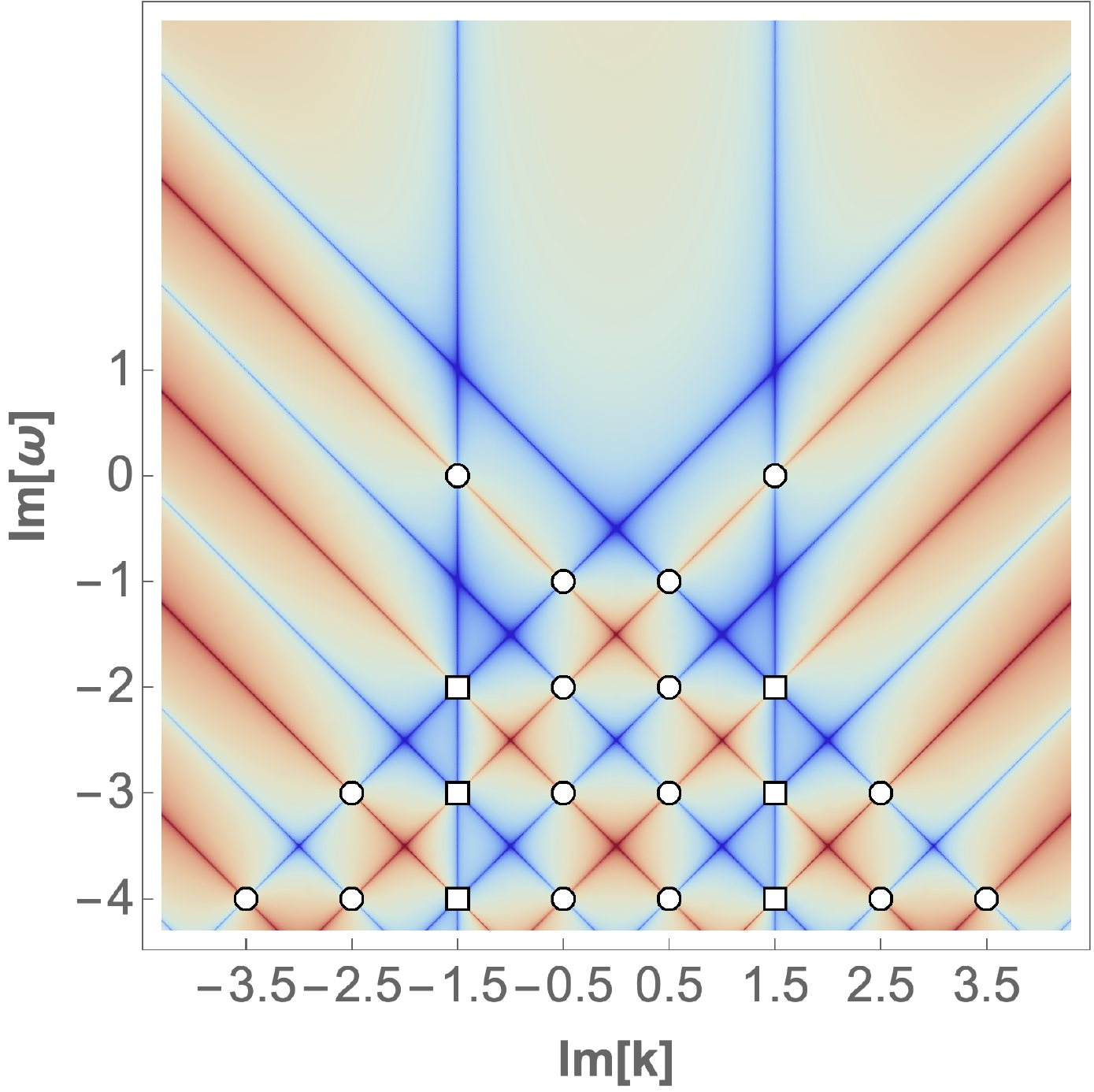}\label{massless123b}}
     \subfigure[$d=6, \Delta=5\,(\delta'=1)$]{\includegraphics[width=4.8cm]{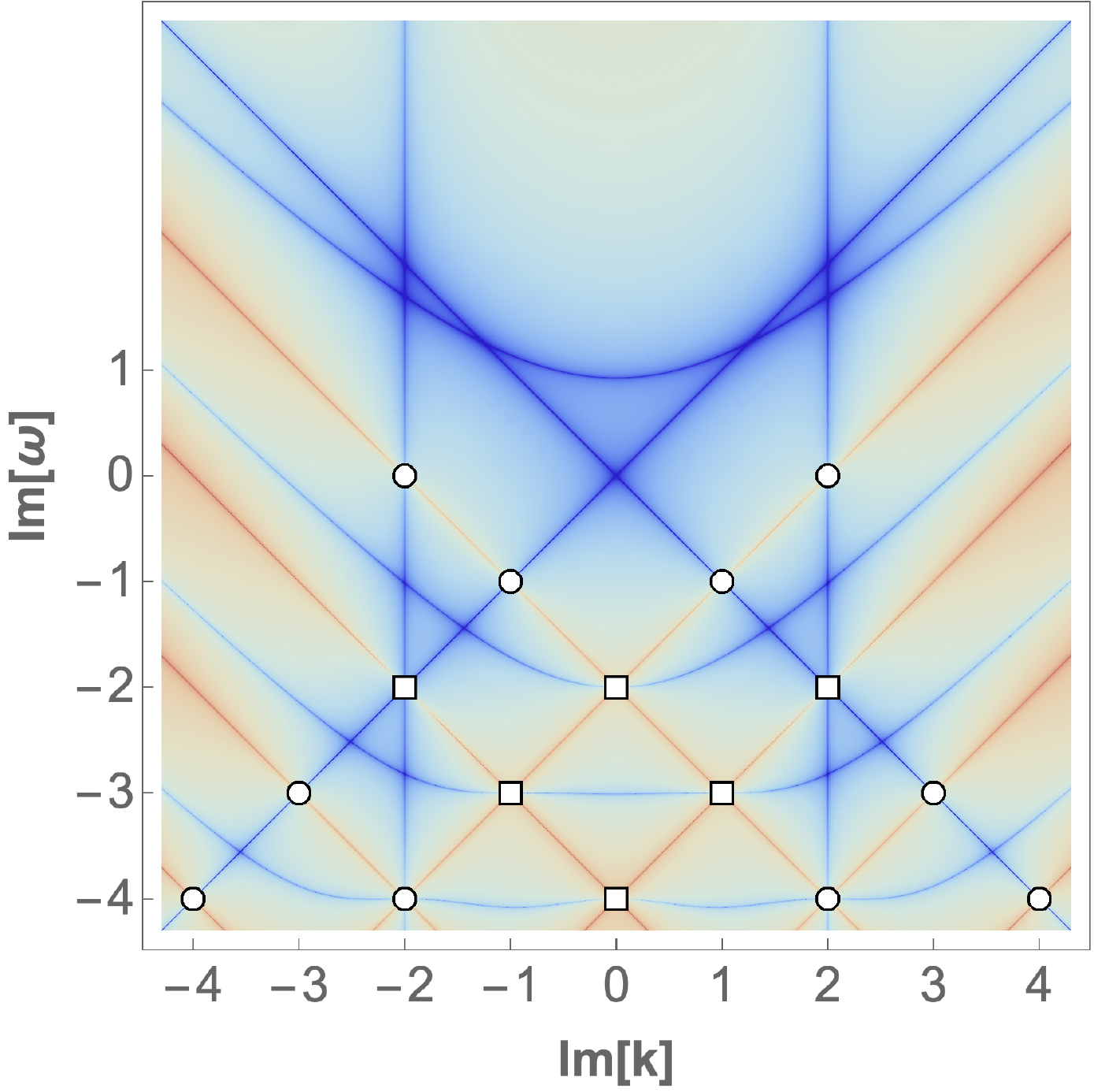}\label{massless123c}}     
     \caption{$\log |\mathcal{G}_V^{\Delta=d-1}(\mathrm{Im}[\omega],\mathrm{Im}[k])|$ at $d=4,5,6$. 
    The blue lines and red lines represent zeros and poles of the Green's function respectively. The white circles and squares are type I and type II pole-skipping points, respectively. The meaning of type I and II will be explained in section \ref{section3}.} \label{massless123}
\end{figure}

\subsubsection{Field theory results}

In this section, we consider a CFT in $S^1\times\mathbb{H}^{d-1}$ and we review the pole-skipping structure of the Green's function for massless vector field perturbations in the ``longitudinal channel'' \cite{Ahn:2020bks}.

  Similarly to the scalar field case \eqref{fourierscalar2pt},\eqref{eq-Gscalar-1}, the Green's function of the vector field depends on the conformal dimension $\Delta$ and on the spacetime dimension $d$. 
  Since $\Delta=d-1$ for the massless case, the Green's function depends only on the spacetime dimension $d$.  
 For $d\geq3$\footnote{The condition $d\geq3$ is natural to make $\mathbb{H}^{d-1}$ a legit hyperbolic space. As the massless vector field always satisfies the unitarity bound $\Delta\geq d-1$, there is no restriction from the unitarity bound.}, the Green's function is
  \begin{equation}
\begin{split}
    \mathcal{G}_V^{\Delta=d-1}(\omega,k)&\propto\frac{\Gamma(x+\delta'/2)\Gamma(y+\delta'/2)}{\Gamma(x-\delta'/2))\Gamma(y-\delta'/2)}\Gamma(-\delta'-1)\\
    &\quad\times\left(k^2+(\delta'+1)^2\right)\,,\label{fouriergauge2pt}
\end{split}
\end{equation}
  for odd $d$ and
\begin{equation}
\begin{split}
    \mathcal{G}_V^{\Delta=d-1}(\omega,k)&\propto\frac{\Gamma(x+\delta'/2)\Gamma(y+\delta'/2)}{\Gamma(x-\delta'/2))\Gamma(y-\delta'/2)}\left[\psi(x+\delta'/2)+\psi(y+\delta'/2)\right]\\
    &\quad\times\left(k^2+(\delta'+1)^2\right)
\,,\label{rvtpf}
\end{split}
\end{equation}  
for even $d$.
Here, 
\begin{equation}
    x=\frac{-i\omega+ik+1}2\,,\quad y=\frac{-i\omega-ik+1}2\,,\quad \delta'\equiv d/2-2 \,.
\end{equation}

The main difference from the scalar case is the existence of additional zero-conditions from the quadratic $k$-term in the second line of \eqref{fouriergauge2pt} and \eqref{rvtpf}. These zero-conditions are represented by the blue vertical lines in Fig. \ref{massless123}, which correspond to $\mathrm{Im}(k)=\pm(\delta'+1)=\pm (d/2-1)$. Aside this pair of vertical blue lines, the procedure to identify the other pole-skipping points is similar to the scalar field's case:

\paragraph{Case 1: non-integer $\delta'$ (odd $d$)} There is no cancellation between the gamma functions in  \eqref{fouriergauge2pt}. This results in an infinite number of zero-lines and pole-lines. See for example Fig. \ref{massless123b} for $\delta'=0.5$.

\paragraph{Case 2: $\delta'\in\mathbb Z^+$ (even $d\neq4$)} The four gamma functions in \eqref{rvtpf}  become a $\delta'$-th order polynomial of $x$ and $y$, which gives conditions for zeros.  For example, if $\delta'=1$ this condition gives two straight blue lines with slopes $\pm 1$ in Fig. \ref{massless123c}. 
The two digamma functions in \eqref{rvtpf} give the condition for poles:
\begin{equation} 
    {-i\omega \pm ik+\delta'+1=-2j,\qquad(j=0,1,2\dots)\,.}
\end{equation}
 This condition corresponds to the straight red lines in Fig. \ref{massless123c}.
Finally, there is the condition for zeros that appears from the sum of the two digamma functions, i.e. 
\begin{equation}
 \psi(x+\delta'/2) + \psi(y + \delta'/2) = 0   \,. 
\end{equation}
They are represented by the blue curves in Fig. \ref{massless123c}.
\paragraph{Case 3: $\delta' = 0$ ($d=4$)} 
Except for the quadratic $k$-term, the Green's function \eqref{rvtpf} is given by the sum of two digamma functions. 
\begin{equation}
 \psi(x) + \psi(y) = 0   \,, 
\end{equation}
from which all the conditions for zeros and poles appear. 
Fig.~\ref{massless123a} shows this case. Note that there are no non-vertical straight blue lines contrary to Fig.~\ref{massless123b} and Fig.~\ref{massless123c}. The blue curves come from the zero-sum condition of the two digamma functions and the red straight lines appear if $x$ and $y$ are non-positive integers. 

At the end of the day, the pole-skipping points in the longitudinal channel of massless vector fields can be summarized as:
\begin{alignat}{3}
& \omega_{0} =0 \ \ &&\mathrm{and} \ \  &&k_{\{0\}} = \pm i\frac{d-2}{2}\,, \\
&\omega_{n} = - in  \ \ &&\mathrm{and} \ \ &&k_{\{n\}} = \pm i\left(-n+2q+\frac{d-6}{2}\right) \,,
\end{alignat}
where $n=1,2,\cdots$ and $q =1,2, \cdots ,n$. The above results agree with the results of the near horizon analysis \eqref{VNHApsps}.  This agreement can be also seen by comparing Fig. \ref{diffpsps} and Fig. \ref{massless123}.

\section{Type-I and type-II pole-skipping points from Green's functions}\label{section3}

It was reported~\cite{Blake:2019otz} that there is an `anomalous' pole-skipping point whose properties are different from usual ones in the sense that the Green's function near the pole-skipping point is not determined by the slope $\delta \omega/\delta k$ as shown in \eqref{jshg74}. 
To our knowledge, the precise meaning and mathematical structure of this anomalous pole-skipping point has not been clarified yet. In this section, we carefully analyse this anomalous point and classify it as a {\it type-II pole-skipping point}, calling the usual ones {\it type-I pole-skipping points}. We also clarify how to distinguish them using a near horizon analysis in the next section.


\subsection{Simple examples and definitions}

\begin{figure}
\centering
    \includegraphics[width=4.9cm]{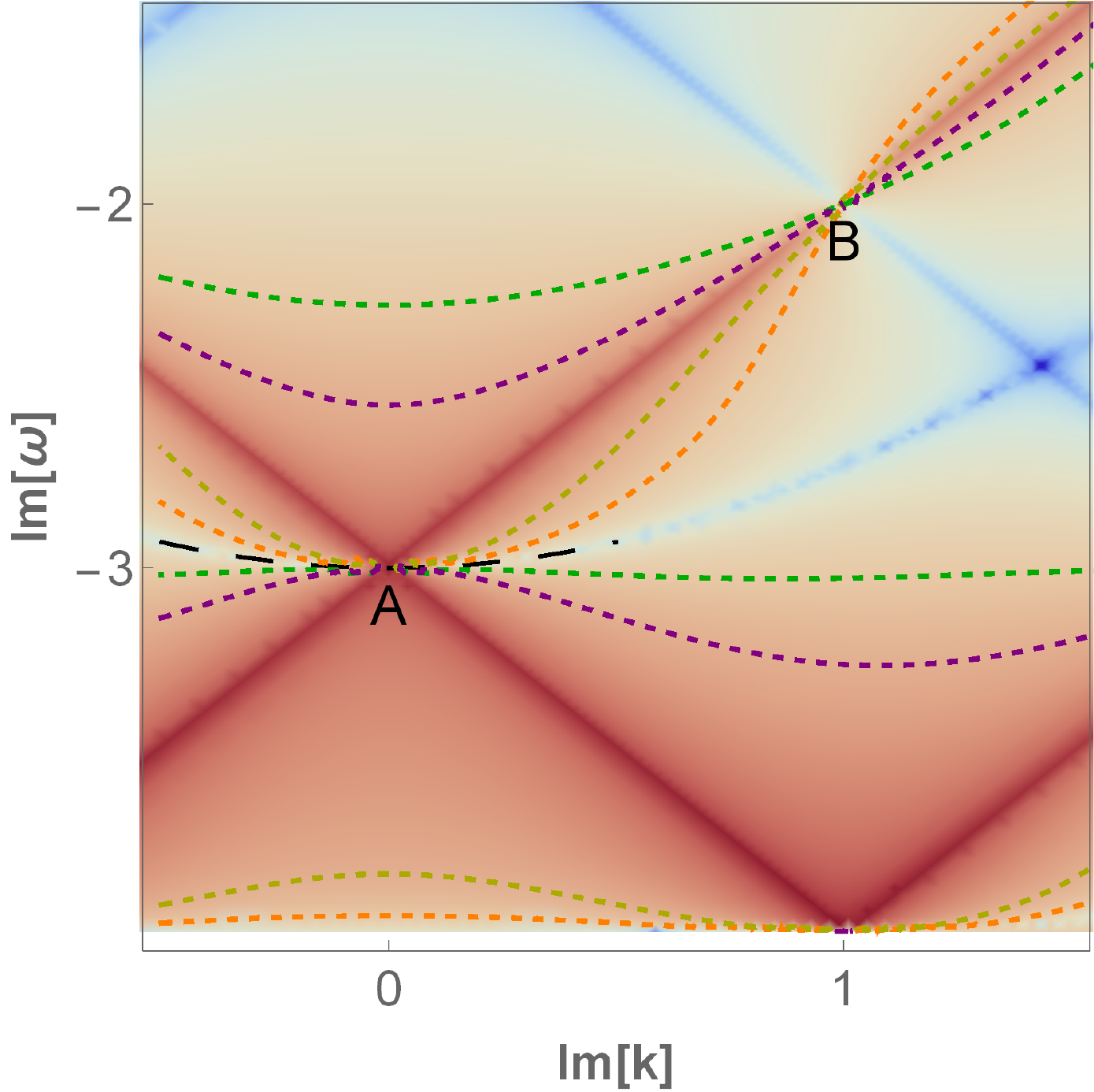}
    \caption{Zoom-in of Fig. \ref{fig:regscalar4} including a type-I point (B) and a type-II point (A). The dotted lines represent curves on which the Green's function has a constant value. Different colors represent different values: $\{\mathrm{black},\mathrm{orange}, \mathrm{yellow}, \mathrm{green}, \mathrm{purple}\} = \{0, 1.05, 3, -1.05, -3\}$. 
   }
    \label{fig:regscalar4magni}
\end{figure}

 We start with simple examples exhibiting the characteristic properties of type-I and type-II points.  Let us consider two points $(0,-3)$ and $(1,-2)$ in Fig. \ref{fig:regscalar4}. 
 They are reproduced as the points A and B in Fig.~\ref{fig:regscalar4magni}, where the dotted curves represent the curves on which the Green's functions take constant values.
 Different colors represent different values: $\{\mathrm{black},\mathrm{orange}, \mathrm{yellow}, \mathrm{green}, \mathrm{purple}\} = \{0, 1.05, 3, -1.05, -3\}$.
From these representative curves, it is clear that there are infinitely many curves of constant values from $-\infty$ to $\infty$, passing through  A and B. Thus, the Green's function is not well-defined both at A and B. In this sense, we might say that both A and B are pole-skipping points. 

A difference between A and B is how the line where the Green's function takes constant values approaches the pole-skipping point. Near B the lines approach with different slopes ($\delta \omega_I/\delta k_I $) while near A the lines approach with a zero slope, $\delta \omega_I/\delta k_I=0$. To find out how the lines approach to the point B, let us investigate the Green's function \eqref{eq-Gscalar-1} near the point A:  
\begin{equation}
\label{twodigamma}
\begin{split}
&\mathcal{G}^\Delta(\omega_{I},k_I)=\mathcal{H}(\omega_I,k_I)\left(\psi \left(\frac{1}{2} (-k_I+\omega_I+3)\right)+\psi\left(\frac{1}{2} (k_I+\omega_I+3)\right)\right)\,,\\ 
&\mathcal{H}(\omega_I,k_I) = \frac{\pi^2 \Gamma \left(\frac{1}{2} (-k_I+\omega_I+3)\right) \Gamma \left(\frac{1}{2} (k_I+\omega_I+3)\right) }{6 \Gamma \left(\frac{1}{2} (-k_I+\omega_I-1)\right) \Gamma \left(\frac{1}{2} (k_I+\omega_I-1)\right)}\,, 
\end{split}
\end{equation}
where $\omega_I:=\Im \omega$ and $k_{I}:= \Im k$. 
In the limit where we approach the point A, we have
%
\begin{align}
    \lim_{\substack{k_{I}\to 0\\ \omega_{I}\to -3}}\mathcal{G}^\Delta(\omega_{I},k_I) &=  \lim_{\substack{k_{I}\to 0\\ \omega_{I}\to -3}}\left\{\psi\left(\frac{1}{2} (-k_{I}+\omega_{I}+3)+1\right)+\psi\left(\frac{1}{2} (k_{I}+\omega_{I}+3)+1\right) \right.\nonumber \\
    &\,\,\left.-\frac{2}{\omega_{I}-k_{I}+3}-\frac{2}{\omega_I + k_{I}+3}\right\}\mathcal{H}(\omega_I,k_I)\nonumber \\
    &=\lim_{\substack{k_{I}\to 0\\ \omega_{I}\to -3}}\left\{-2\gamma-\frac{4(\omega_{I}+3)}{(\omega_{I}+3)^2-k_{I}^2}\right\}\frac{\pi^2}{96} \\
    &=\lim_{\substack{k_{I}\to 0\\ \tilde{\omega_{I}}\to 0}} \frac{2\gamma k_I^2-4 \tilde{\omega}_{I}}{\tilde{\omega}_{I}^2-k_{I}^2}\frac{\pi^2}{96}  \qquad (\tilde{\omega}_I := \omega_I + 3) \label{twodigammaprime} \,,
\end{align}
where we used $\psi(x) = \psi(x+1)-\frac{1}{x}$ and $\psi(1)=-\gamma$.\footnote{$\gamma \approx 0.577$ denotes the Euler-Mascheroni constant.} We dropped the term of order $\tilde{\omega}_I^2$ in the numerator in \eqref{twodigammaprime} because we are taking a small $\tilde{\omega}_I$ limit.  

The limiting value in \eqref{twodigammaprime}  is not unique and depends on the path in which one approaches the pole-skipping point $(\tilde{\omega}_I, k_I) = (0,0)$. If we choose a linear path $\tilde{\omega}_I = q_1 k_I$, the Green's function diverges. However, if we consider a quadratic path $\tilde{\omega}_I = q_2 k_I^2$, the Green's function is finite and depends on the `curvature' $q_2$:
\begin{equation} \label{lhybg}
     \lim_{\substack{k_{I}\to 0}}\mathcal{G}^\Delta(\tilde{\omega}_{I},k_I) =\frac{\pi^2}{96}\left( - 2\gamma +4 q_2\right) \,.
\end{equation}
For example, along a line of zeros (blue line) \eqref{lhybg} vanishes, so $q_2 = \gamma/2$. We confirmed that this is indeed the case:  the black dashed curve in Fig. \ref{fig:regscalar4magni} is given by $\tilde{\omega}_I = q_2\, k_I^2= (\gamma/2) k_I^2$, which overlaps the blue curve around A.
We also confirmed that the other dotted curves 
$\{\mathrm{orange}, \mathrm{yellow}, \mathrm{green}, \mathrm{purple}\} = \{ 1.05, 3, -1.05, -3\}$
are consistent with \eqref{lhybg}, i.e. $q_2= \{ 24\pi^2 (1.05) + 1/2\gamma, 24\pi^2 (3) + 1/2\gamma, 24\pi^2 (-1.05)+ 1/2\gamma, 24\pi^2 (-3) + 1/2\gamma  \} $ respectively.  In addition, the two lines of poles passing through A are explained by the zeros of the denominator in \eqref{twodigammaprime}.

Based on this observation, we define the type-I and type-II pole-skipping points as follows. If {\it all} the contour curves of the Green's function approach the pole-skipping point linearly, we call it type-I. Otherwise, we call it type-II. In the above example, the contour curves in which the Green's function is finite approach the type-II point quadratically. More generally speaking, in the type-II case, the lines with constant values can be any higher order ($\omega = q_n k^n,  n\ge2$) than linear and they are tangential to a line of zeros or to a line of poles.  To distinguish the type-I and type-II points we use white circles for type-I and white squares for type-II in Figs.~\ref{fig:regscalar} and  \ref{massless123}.  

%

\subsection{General forms of the Green's functions}
\label{sec:GFOG}

\paragraph{More examples} 
The type-II points appear in both scalar and massless vector cases and they are marked by the white squares in Figs.~\ref{fig:regscalar4}, \ref{fig:regscalar2.0}, and \ref{massless123}. All cases except Fig.~\ref{massless123b} are similar, involving two pole-lines and one curve of zeros, which comes the sum of two digamma functions, as in \eqref{twodigammaprime}.

Fig.~\ref{massless123b} shows another pattern for type-II points, which involves the intersection of two lines of zeros and one line of poles. Contrary to the previous case, if we approach the pole-skipping point linearly, the Green's functions vanishes instead of diverging.

There is yet another pattern for pole-skipping points in  Fig.~\ref{massless123a}, in which {\it two} lines of zeros and {\it two} lines of poles intersect. It turns out that this is also a type-I point. 

\paragraph{General Green's function near the type-I and type-II points} 
Built on the above observations, we develop a general criteria to distinguish type-I and type-II points in terms of the general form of the Green's function. 
Let us consider a retarded Green's function of the form
\begin{equation}
    G^R(\omega, k) \sim \frac{\mathcal{B}(\omega , k)}{\mathcal{A}(\omega ,k)},\label{dgr}
\end{equation}
where $\mathcal{A}(\omega , k)$ and $\mathcal{B}(\omega ,k)$ are functions which are related to poles and zeros of the retarded Green's function. 

For any pole skipping-point $(\omega_*, k_*)$, we may shift the momentum $(\omega, k) \rightarrow (\omega-\omega_*,k-k_*)$ so that the pole-skipping point is located at $(\omega_*, k_*)=(0,0)$. See for example \eqref{twodigammaprime}. 
 In general, $\mathcal{A}(\omega , k)$ and $\mathcal{B}(\omega ,k)$ can be expanded around the pole-skipping point $(0,0)$:
\begin{equation}
\label{taylor}
\begin{split}
    &G^R( \delta \omega,  \delta k)   \sim \frac{\mathcal{B}(\delta \omega, \delta k)}{\mathcal{A}(\delta \omega, \delta k)}\\ & = \frac{\mathcal{B}|_* +  \left(\delta \omega \partial_{\omega}\mathcal{B} + \delta k \partial_k \mathcal{B}\right)|_* + \frac{1}{2} \left( (\delta \omega)^2 \partial^2_{\omega}\mathcal{B} + (\delta k)^2 \partial^2_k \mathcal{B} + 2 (\delta \omega \delta k) \partial_\omega \partial_k  \mathcal{B} \right)|_*  +\cdots}{\mathcal{A}|_* +  \left(\delta \omega \partial_{\omega}\mathcal{A} + \delta k \partial_k \mathcal{A}\right)|_* + \frac{1}{2} \left( (\delta \omega)^2 \partial^2_{\omega}\mathcal{A} + (\delta k)^2 \partial^2_k \mathcal{A} + 2 (\delta \omega \delta k) \partial_\omega \partial_k  \mathcal{A} \right)|_*  +\cdots} \,.
\end{split}
\end{equation}
where $|_*$ means $|_{(\omega, k)=(0,0)}$. 
At the pole-skipping point ($\delta \omega = \delta k =0$), $\mathcal{A}|_*=\mathcal{B}|_* = 0$ so 
the retarded Green's function is ill-defined:
\begin{equation}
    G^R(\omega_*,k_*) \sim \frac{0}{0}\,.
\end{equation}
However, the Green's function can be well defined near the pole-skipping point, if $\delta \omega \ne 0$ and/or $\delta k \ne 0$. 

It is possible that the Green's function near the pole-skipping point is determined by the `slope' $\delta \omega/\delta k$~\cite{Blake:2018leo, Blake:2019otz, Natsuume:2019sfp, Natsuume:2019xcy, Natsuume:2019vcv} However, if some of the first derivatives in \eqref{taylor} vanish this may not be the case.
 Here, we consider the most general case that the first non-vanishing orders of $\mathcal{B}$ and $\mathcal{A}$ in \eqref{taylor} are $N$ and $M$, respectively. In this case, the retarded Green's function should have the following structure for small enough  $(\delta \omega, \delta k)$, 
\begin{equation} \label{ojegsa}
    G^R(\delta \omega , \delta k) \sim \frac{\prod_{i=1}^N(z_{1,i}\delta \omega-z_{2,i} \delta k)}{\prod_{j=1}^M(p_{1,j}\delta \omega-p_{2,j} \delta k)},
\end{equation}
where $N,M \in \mathbb{N}$. The coefficients $z_{1,i}, z_{2,i}, p_{1,j}, p_{2,j}$ are some number. We further assume that there is no common factor in the numerator and denominator. If there is a common factor there is a subtlety, which will be dealt with separately.
The slope dependence of the Green's function is revealed by taking a limit along the path $\delta \omega = q_1 \delta k$
\begin{equation}
\label{dettype1}
    \lim_{\delta k \rightarrow 0}G^R(q_1\delta k , \delta k) \sim 
    \lim_{\delta k \rightarrow 0}\delta k^{N-M}\frac{\prod_{i=1}^N(z_{1,i}q_1-z_{2,i})}{\prod_{j=1}^M(p_{1,j}q_1-p_{2,j})}\,.
\end{equation}

The equation \eqref{dettype1} gives us a classification of the pole-skipping points.  
\begin{itemize}
\item
If $N = M$ (the numbers of factors in the denominator and numerator are the same) then (\ref{dettype1}) depends on $q_1$. This defines a type-I pole-skipping point.   
\item If $N\neq M$ (the numbers of factors in the denominator and numerator are different) then (\ref{dettype1}) is zero or infinite so it does not depend on $q_1$. This defines a type-II pole-skipping point. 
\item
Graphically speaking, if the number of zero-lines and the number of pole-lines intersecting at the pole-skipping point are the same (different), then the point is  type-I (type-II). All our type-I points in Figs.~\ref{fig:regscalar} and \ref{massless123}; and all our type-II points in Figs.~\ref{fig:regscalar} and \ref{massless123} follow this rule. 
\end{itemize}

There may be a special case, where there is a common factor in the numerator and the denominator of \eqref{ojegsa}. If a general form \eqref{ojegsa} remains valid after a cancellation of the common factor due to other remaining factors, nothing changes in our conclusion. However, if the form \eqref{ojegsa} just becomes a constant because of the cancellation, then we have a type-II point even though $M=N$. Let us explain this special case with an example. Suppose the Green's function is of the form 
\begin{equation} \label{cgshe}
G^R_2(\omega, k) = \frac{\omega- k^2}{\omega} \,.
\end{equation} 
Here, $(\omega, k)=(0,0)$ is a type-II pole-skipping point because near this point
\begin{equation}
G^R_2(\delta \omega, \delta k) = \frac{\delta \omega}{\delta \omega} = 1  \,.
\end{equation} 
which does not depend on $q_1$ 
along the linear path $\delta \omega = q_1 \delta k$. In this case, we may include higher-order terms, like $(\delta k)^2$. Then, for a  quadratic path $\delta \omega = q_2\delta k^2$ 
\begin{equation} \label{kgsber}
\lim_{\delta k \rightarrow 0}G^R_2(q_2\delta k^2 , \delta k) \sim\frac{q_2-1}{q_2}.
\end{equation}
 Graphically speaking, even if the number of zero-lines and the number of pole-lines are the same, the point can be of type-II if the lines are tangential to each other at the pole-skipping point.

\paragraph{Physical constraints and path-dependence}
The above analysis describes a purely  mathematical structure of the Green's function near the pole-skipping point. If we consider our gravity system in the context of holography, the form of \eqref{ojegsa} is more constrained thanks to the properties of the black hole horizon. For example, the type-I point will always have the simple form of \eqref{jshg7} ($N=M=1$)\footnote{This statement is only true for the case without a prefactor such as the scalar case. With a prefactor, the type may be changed, see Sec.~\ref{Sec5}, For example, Fig. \ref{massless123a} has Type-I with $(N=M=2)$, i.e. the intersection of two zero lines and two pole lines.}  and the type-II point will have one of the simple forms given in \eqref{jshg74}, which are at most of quadratic order in $(\delta \omega, \delta k)$ ($N$ and $M$ are not greater than 2).

Another important physical constraint comes from the non-uniqueness of incoming boundary condition, which implies that the $i\omega$ at the pole-skipping point is integer for bosons and half integer for fermions. However, there is an exceptional case with non-integer (or half integer) $i\omega$ pole-skipping point. We call it type-III and in this case, a UV condition is also involved. 

If we just want to judge whether the pole-skipping point is of type-I or type-II it is enough to retain the leading power in every factor as in \eqref{ojegsa}. However, in the type-II case, if we want to specify the path at which the Green's function is constant we have to retain the second order term in every factor as in \eqref{kgsber} and \eqref{jshg74}. For example, let us come back to the type-II point  \eqref{twodigammaprime}. This is the case with $N=1, M=2$ in \eqref{ojegsa} and the second case in \eqref{jshg74}.  If we want to figure out the path at which the Green's function is constant we need to keep terms up to quadratic order.

\section{Identifying type-I and type-II from the near horizon analysis} \label{Sec4}

In section~\ref{SNH}, we obtained the pole-skipping points performing a near horizon analysis. 
In this section, we ask if the near horizon analysis  can also identify the type of the pole-skipping points. The answer is almost yes, but there is a small subtlety. 

Classifying the different types of pole-skipping points is related to the behavior of the Green's function near them. Thus, we need to do the near horizon analysis in the vicinity of the pole-skipping points. The formalism to consider points nearby the pole-skipping points has been already developed in Sec.~\ref{scalar11}. We may use all the formulas there but with a different interpretation. For example, let us consider \eqref{lmnbg7} 
\begin{equation} \label{lmnbg77}
    \phi(z;\omega,k) = \phi_0(\omega,k) + \cdots +  (1-z)^n (\phi_n(\omega,k) + \phi_{n+1}(\omega,k)(1-z) + \cdots) \,,
\end{equation}
where we recovered the arguments $(\omega, k)$ to emphasize that the equation can be applied to non-pole-skipping points as well as pole-skipping points. Here, because we consider non-pole-skipping points, $\phi_n$ is a function of $\phi_0$. If this were about the pole-skipping point, $\phi_n$ would have been independent of $\phi_0$. 

In the vicinity of the pole-skipping points the Green's function is a function of $\phi_n/\phi_0$. Thus, one may think that if $\phi_n/\phi_0$ depends on $\delta \omega/\delta k$ near the pole-skipping point then the point is of type-I. This is almost true but there is a subtlety. 

The analysis of Sec.~\ref{sec41} and \ref{sec42} has been essentially done in ~\cite{Blake:2019otz}. Here, we first review it in the context of our model to setup the stage. Next, we explain the relation between the general structure of the Green's function in Sec.~\ref{sec:GFOG} and the near horizon analysis. In this section, we focus on the scalar case. In Sec.~\ref{Sec5}, we deal with a subtlety  that appears in the vector and, in general, higher spin cases.

\subsection{Simple examples} \label{sec41}

For example, let us consider two points $(-2,-1)$ in Fig.~\ref{fig:regscalar4} and $(0,-1)$ in Fig.~\ref{fig:regscalar2.0}. The former is a type-I point and the latter is a type-II point. 
Because they are the highest pole-skipping points it is enough to consider the $\mathcal{O}((1-z)^0)$ equation \eqref{dksgs}.

The type-I point $(-2,-1)$ in Fig.~\ref{fig:regscalar4} corresponds to $d=4$ and $\Delta = 4$.  Near the pole-skipping point $\omega = -i(1+\delta \omega)$ and $k= i(-2+\delta k) $, \eqref{dksgs} must be satisfied, which implies
\begin{equation}
    \left(-4\delta k + 3  \delta \omega \right)\phi_0 + 2  \delta \omega \phi_1=0.
\end{equation}
Along the path $\delta \omega = q_1 \delta k$, the ratio $\phi_1/\phi_0$ in the small $\delta k$ limit is
\begin{equation}
\label{type2scar}
    \frac{\phi_1}{\phi_0}=-\frac{3}{2}+\frac{2}{q_1}\,,
\end{equation}
which shows that indeed $\frac{\phi_1}{\phi_0}$ depends on the slope $q_1=\delta \omega/\delta k$.

The type-II point $(0,-1)$ in Fig.~\ref{fig:regscalar2.0} corresponds to $d=4$ and $\Delta = 2$.  Near the pole-skipping point $\omega = -i(1+\delta \omega)$ and $k= 0+ \delta k $, \eqref{dksgs} must be satisfied, which implies
\begin{equation}
\label{ex1scar}
  ((\delta k)^2 + 3  \delta \omega ) \phi_0 +2   \delta \omega \phi_1=0\,.
\end{equation}
Along the path $\delta \omega = q_1 \delta k$, the ratio $\phi_1/\phi_0$ in the small $\delta k$ limit is
\begin{equation}
\label{type1scar}
    \frac{\phi_1}{\phi_0}=-\frac{3}{2}\,,
\end{equation}
which does not depend on $q_1$.  However, if we consider a quadratic path $\delta \omega = q_2 (\delta k)^2$, then
\begin{equation}
    \frac{\phi_1}{\phi_0}=-\frac{1-3q_2}{2q_2}\,,
\end{equation}
which shows that $\frac{\phi_1}{\phi_0}$ depends on $q_2$, i.e., it depends on the particular quadratic curve along which we approach the pole-skipping point.

In next subsection, we discuss concrete relations between the near horizon analysis and the Green's function, as well as their general relation. 

\subsection{Relation between near horizon analysis and the Green's function} \label{sec42}

The relation \eqref{pspeom1}
\begin{equation}
    \label{pspeom11}
    \det \mathcal{M}^{(n)}(\omega,k)\phi_0(\omega, k) + \mathcal{N}^{(n)}(\omega)(n - i \omega) \phi_n(\omega, k)=0 \,,
\end{equation}
is valid for any $(\omega, k)$ regardless of whether it is a pole-skipping point or not. At a pole-skipping point, ($\omega_n, k_j$), $\phi_0$ and $\phi_n$ are independent. However, in the vicinity of a pole-skipping point,
\begin{equation}
    \omega = \omega_n + \delta \omega\,, \quad k = k_{j} + \delta k\,,
\end{equation}
%
%
 the ratio between two coefficients $\phi_0$ and $\phi_n$ is fixed by $\delta \omega, \delta k$:
\begin{equation} \label{gscqg}
   \frac{\phi_n(\omega_n+\delta \omega,k_{j} + \delta k)}{\phi_0(\omega_n+\delta \omega,k_{j} + \delta k)} = \frac{ \det \mathcal{M}^{(n)}(\omega_n+\delta \omega,k_{j} + \delta k)}{i \delta \omega  \mathcal{N}^{(n)}(\omega_n + \delta \omega)} \,.
\end{equation}
Having the information \eqref{gscqg} for points near the pole-skipping point, we now want to ask the following question: 
near a pole-skipping point, how to identify the path on which the Green's function is constant.

For convenience, let us shift the coordinate $\omega \to \omega -\omega_n$ and $k \to k-k_j$. Then, \eqref{gscqg} yields
\begin{equation} \label{sgdce}
   \frac{\phi_n|_*}{\phi_0|_*} \equiv \frac{\phi_n(0,0)}{\phi_0(0,0)} \equiv  \lim_{\delta \omega, \delta k \to 0} \frac{ \det \mathcal{M}^{(n)}(\delta \omega, \delta k)}{i \delta \omega  \mathcal{N}^{(n)}( \delta \omega)} \,.
\end{equation}
We emphasize that the double limit in the second relation provides a path-dependent definition for $(\phi_n/\phi_0)|_*$, i.e, the value it takes depends on the functional relation between $\delta \omega$ and $\delta k$ in the curve that we use to take the limit. This is just another way of saying that the ratio is not well-defined at the pole-skipping point.

Next, we want to connect the ratio $(\phi_n/\phi_0)|_*$ to the Green's function, which means the relation between IR ($z\sim 1$) and UV ($z \sim 0$) quantities. Let us start with the UV limit, in which the field can be expressed as, at any $(\omega, k)$,
\begin{equation}
\label{ansatz:incomingsol}
    \phi\left(z;\omega, k\right) =  \phi^{[\mathbf{nn}]}(z;\omega,k) - \mathcal{G}\left( \omega, k \right) \phi^{[\mathbf{n}]}(z;\omega,k) \,,
\end{equation}
where $\mathcal{G}$ is proportional to the Green's function and the `$-$' sign is introduced for later convenience.  $\phi^{[\mathbf{nn}]}$ is a non-normalizable mode and $\phi^{[\mathbf{n}]}$ is a normalizable mode. For example, in the case of a scalar field,
\begin{equation}
   \phi^{[\mathbf{nn}]}(z;\omega,k) =  z^{d-\Delta} + \cdots \,,  \qquad \phi^{[\mathbf{n}]}(z;\omega,k) = z^{\Delta} +\cdots  \,.
\end{equation}

At IR and at a pole-skipping point, \eqref{ansatz:incomingsol} can be expanded as 
\begin{equation}
\label{}
    \phi(r)|_* \equiv \sum_{i=0} \phi_i|_* (1-z)^i  = \sum_{i=0}\left. \left(\phi^{[\mathbf{nn}]}_i  - \mathcal{G} \phi^{[\mathbf{n}]}_i \right) \right|_* (1-z)^i \,.
\end{equation}
Note that at a {\it nearby point} of a pole-skipping point, there is another independent singular solution  such as
$  (1-z)^{i\omega} \sum_{i=0}\tilde\phi_i(1-z)^i$ \footnote{There may be an additional $\log (1-z)$ term if the nearby points are on the line of integer $i\omega$. } . In this case,   $\mathcal{G} = -\tilde{\phi}^{[\mathbf{nn}]}_0/\tilde{\phi}^{[\mathbf{n}]}_0$ is uniquely determined to keep regularity. However, at the pole-skipping point with integer $i\omega$, we do not have that regularity condition so $\mathcal{G}$ can be freely chosen. In terms of the boundary condition at the horizon $z=1$, it amounts to determining the ratio $\phi_n/\phi_0$ freely as follows.  
\begin{equation} \label{zsdfgw}
   \left. \frac{\phi_n}{\phi_0} \right|_* =  \left.
   \frac{\phi^{[\mathbf{nn}]}_n  - \mathcal{G} \phi^{[\mathbf{n}]}_n }{  \phi^{[\mathbf{nn}]}_0  - \mathcal{G} \phi^{[\mathbf{n}]}_0 } \right|_*    \,.
\end{equation}
If $\mathcal{G}=0(\infty)$ the ratio $ (\phi_n/\phi_0)|_*$ is determined by the non-normalizable (normalizable) mode as it must be.

By combining \eqref{sgdce} and \eqref{zsdfgw}, we may define the Green's function at the pole-skipping point:
\begin{equation}
\label{eq:GRconnection}
    \mathcal{G}|_*  \equiv   \lim_{\delta \omega, \delta k \to 0} \frac{\det \mathcal{M}^{(n)}\left(\delta \omega, \delta k\right)  \phi^{[\mathbf{nn}]}_0|_*-i \delta \omega  \mathcal{N}^{(n)} (\delta \omega) \phi^{[\mathbf{nn}]}_n|_*}{\det \mathcal{M}^{(n)}\left(\delta \omega, \delta k\right)  \phi^{[\mathbf{n}]}_0|_*-i \delta \omega  \mathcal{N}^{(n)}(\delta \omega) \phi^{[\mathbf{n}]}_n|_*}  \equiv \lim_{\delta \omega, \delta k \to 0}  \frac{\mathcal{B}(\delta \omega, \delta k)}{\mathcal{A}(\delta \omega, \delta k)}\,.
\end{equation}
where in the second equality we define $\mathcal{B}(\delta \omega, \delta k)$ and $\mathcal{A}(\delta \omega, \delta k)$ which  can be identified with the ones in \eqref{taylor}.
By this identification, we find that $\mathcal{A}$ and $\mathcal{B}$ are linear combinations of the same objects $\det \mathcal{M}^{(n)}$ and $\mathcal{N}^{(n)}$. It gives strong  constrains to the structure of the Green's function as follows.

For example, at the leading order 
\begin{equation}
\label{taylor1}
    G^R( \delta \omega,  \delta k)   \sim \frac{\mathcal{B}(\delta \omega, \delta k)}{\mathcal{A}(\delta \omega, \delta k)}= \frac{ \left(\delta \omega \partial_{\omega}\mathcal{B} + \delta k \partial_k \mathcal{B}\right)|_*}{  \left(\delta \omega \partial_{\omega}\mathcal{A} + \delta k \partial_k \mathcal{A}\right)|_* } \,.
\end{equation}
where
\begin{align}
\partial_\omega \mathcal{B} &=  \partial_\omega  \det \mathcal{M}^{(n)} \phi^{[\mathbf{nn}]}_0|_* -i \mathcal{N}^{(n)} \phi^{[\mathbf{nn}]}_n|_*  \label{kghby1} \\
\partial_\omega \mathcal{A} &=  \partial_\omega  \det \mathcal{M}^{(n)} \phi^{[\mathbf{n}]}_0|_* -i \mathcal{N}^{(n)} \phi^{[\mathbf{n}]}_n|_*  \label{kghby2} \\
\partial_k \mathcal{B} &=  \partial_k \det \mathcal{M}^{(n)} \phi^{[\mathbf{nn}]}_0|_* \\
\partial_k \mathcal{A} &=  \partial_k \det \mathcal{M}^{(n)} \phi^{[\mathbf{n}]}_0|_*  \,.
\end{align}
If $\partial_k \det \mathcal{M}^{(n)}|_* \ne 0$, both $\partial_k \mathcal{A}$ and $\partial_k \mathcal{B}$ are nonzero. From the structure of \eqref{kghby1} and \eqref{kghby2}, $\partial_\omega \mathcal{A} $ and $\partial_\omega \mathcal{B} $ cannot be zero simultaneously. Therefore, in this case the pole-skipping point is guaranteed to be type-I at the intersection of only one lines of zeros and one line of poles. The schematic Green's function near the pole-skipping point is
\begin{equation} \label{jshg7}
   \frac{c_1 \delta \omega + c_2 \delta k}{ c_3 \delta \omega + c_4 \delta k} \,,  
\end{equation}
where $c_1$ or $c_3$ can be zero while $c_2 \ne 0$ and $c_4 \ne 0$. 

If $\partial_k \det \mathcal{M}^{(n)}|_* = 0$, both $\partial_k \mathcal{A}$ and $\partial_k \mathcal{B}$ are zero so in this case the pole-skipping point is of  type-II, regardless of  $\partial_k \mathcal{A}$ and $\partial_k \mathcal{B}$. If $\partial_k^2 \det \mathcal{M}^{(n)}|_* \ne 0$, there are three possibilities for the form of the Green's function
\begin{equation} \label{jshg74}
    \frac{c_1 \delta \omega + c_2 (\delta k)^2}{ c_3 \delta \omega + c_4 (\delta k)^2} \,,  \quad   \frac{c_1 \delta \omega + c_2 (\delta k)^2}{ c_3 (\delta \omega)^2 + c_4 (\delta k)^2 + c_5 \delta \omega \delta k} \,, \quad  \frac{c_1 (\delta \omega)^2 + c_2 (\delta k)^2 + c_5 \delta \omega \delta k }{ c_3 \delta \omega + c_4 (\delta k)^2 } \,,
\end{equation}
where at least one $\delta \omega$ has to be retained because $\partial_\omega \mathcal{A}$ and $\partial_\omega \mathcal{B}$ can not vanish simultaneously. 

We now can answer our question: near a pole-skipping point, how to identify the path on which the Green's function is constant from the information in near horizon analysis. If $\partial_k \det \mathcal{M}^{(n)}|_* \ne 0$ (type I), as we see in \eqref{jshg7} we may determine the linear path $\delta \omega = q_1 \delta k$, where $q_1$ gives the Green's function. If $\partial_k \det \mathcal{M}^{(n)}|_* = 0$ and $\partial_k^2 \det \mathcal{M}^{(n)}|_* \ne 0$ (type II), as we see in \eqref{jshg74} we may determine the quadratic path $\delta \omega = q_2 \delta k^2$, where $q_2$ gives the Green's function. All examples in Fig.~\ref{fig:regscalar}  belong to \eqref{jshg7} and \eqref{jshg74}.  These observations are essentially made in \cite{Blake:2019otz} and \cite{Natsuume:2019vcv}. Here, i) we shift the focus to the meaning of the path on which the Green's function is constant; ii) we will show this may not be the whole story when we consider the vector field.  



\subsection{Merging of pole-skipping points and type-II} \label{sec43}
We may understand the type-II point from the merging of two pole-skipping points. We explain it by three methods: by figures, by algebraic formulas of pole-skipping points, and by near horizon analysis. 

\paragraph{By figures} For example, we describe it by comparing Figures in Fig. \ref{fig:regscalar}.
Let us start with Fig. \ref{fig:regscalar4.5}, where $\delta = 2.5$. As $\delta$ decrease the blue lines come close to the red lines: see Fig. \ref{fig:regscalar4.2}, where $\delta=2.2$. When  $\delta$ decreases even further and becomes an integer, $\delta = 2$, we obtain Fig. \ref{fig:regscalar4}, where the blue lines approaching to the red lines disappear and are replaced by the blue curves. The disappearance of the blue lines can be understood by \eqref{con100} and the appearance of the blue curves can be seen by \eqref{con300}.  After that, when $\delta = 1.8$, the blue lines appear again below the red lines. What are the consequences of these movements of the red lines and blue lines/curves in terms of  the pole-skipping points? Note that there are three pairs of pole-skipping points near $(0,-3)$ and $(\pm1,-4)$ at the bottom of Fig. \ref{fig:regscalar4.2} and \ref{fig:regscalar3.8}. 
At $\delta=2$ every pair merges into one point. These merged points correspond to type-II points. All the others points are type-I points. In Fig. \ref{fig:regscalar4.2} and \ref{fig:regscalar3.8} there are only type-I points. All points in Fig.~\ref{fig:regscalar2.0} are type-II points, where pairs of two pole-skipping points will appear if we slightly deviate from $\delta=2$.  

\paragraph{By formulas} What is the algebraic formula corresponding to the type-II point and, more generally, to the case with $\delta \in \mathbb{Z}^*$? It is still \eqref{sps} even though it was originally obtained for $\delta \notin \mathbb{Z}^*$ by imposing non-negative integer condition to the arguments of the gamma functions in \eqref{fourierscalar2pt}.   For example, let us consider $n=3$ and $n=4$. From \eqref{sps} if $n=3$
\begin{equation} \label{lmjbg}
    \mathrm{Im}[k] = \pm (2q-2) = \pm0(q=1), \, \pm2(q=2), \, \pm4(q=3)\,, 
\end{equation}
and if $n=4$
\begin{equation} \label{lmjbg1}
\mathrm{Im}[k] = \pm (2q-3) = \mp1(q=1), \, \pm1(q=2), \, \pm3(q=3), \, \pm5(q=4)    \,.
\end{equation}
Note that the `repeated' points $\pm0(q=1)$ in \eqref{lmjbg} and $\mp1(q=1), \, \pm1(q=2),$ in  \eqref{lmjbg1} correspond to the `merging' of points explained in the above paragraph.

\paragraph{By near horizon analysis}
In general, $\det \mathcal{M}^{(n)}$ has the following structure 
\begin{equation}
\label{pspeom123}
\det \mathcal{M}^{(n)} = \alpha \left[\prod_{i=1}^{2n}(k -  k_{i})+\left(\omega-\omega_n\right)f(\omega, k)\right] \,,
\end{equation}
where $\alpha$ is a numerical constant and  $f(\omega, k)$ is some function. This can be understood by the fact that $\det \mathcal{M}^{(n)}$ is a $2n$ order equation, whose solutions are the pole-skipping points $(\omega_n, k_i \in k_{\{n\}})$. For some values of $d$ and $\Delta$ some of the $k_i$'s can be the same. In other words, some of the pole-skipping points can be multiple solutions of the the $2n$ order equation, $\det \mathcal{M}^{(n)} =0$. This explains the `merging' of pole-skipping points. In this case $\partial_k\det \mathcal{M}^{(n)}|_* = 0$, which yields a type-II pole-skipping point.

\section{Subtleties of near horizon analysis and type-III pole-skipping point} \label{Sec5}

Using the example of a scalar field, we have shown that we may find the pole-skipping points and determine its type only by the near horizon analysis without computing the full Green's function. This is not only practically helpful but also conceptually very important, since it means the existence and properties of the pole-skipping points are originated from the properties of the black hole horizon.

To be sure about this, we compare types of pole-skipping points from two methods: direct  Green's function computation and near horizon analysis. Our goal is to check if the near horizon analysis can fully capture the existence and properties of pole-skipping points of the Green's function. We find that indeed there are two subtleties.

\begin{figure}
\centering
    \subfigure[$d=4, \Delta = 4.2$]{\includegraphics[width=4.8cm]{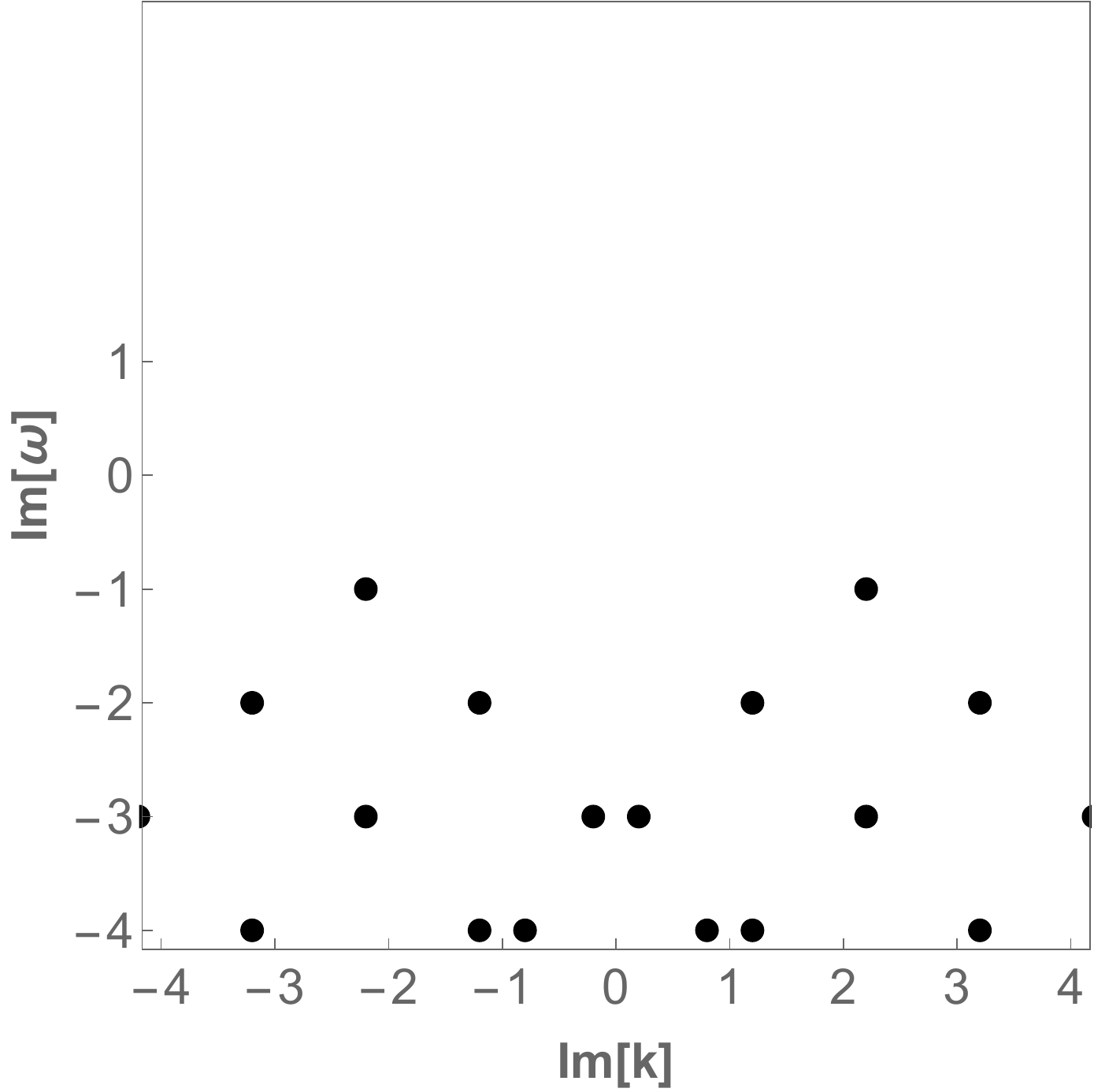}
    \label{holscalpspa}
    }
    \subfigure[$d=4, \Delta = 4$]{\includegraphics[width=4.8cm]{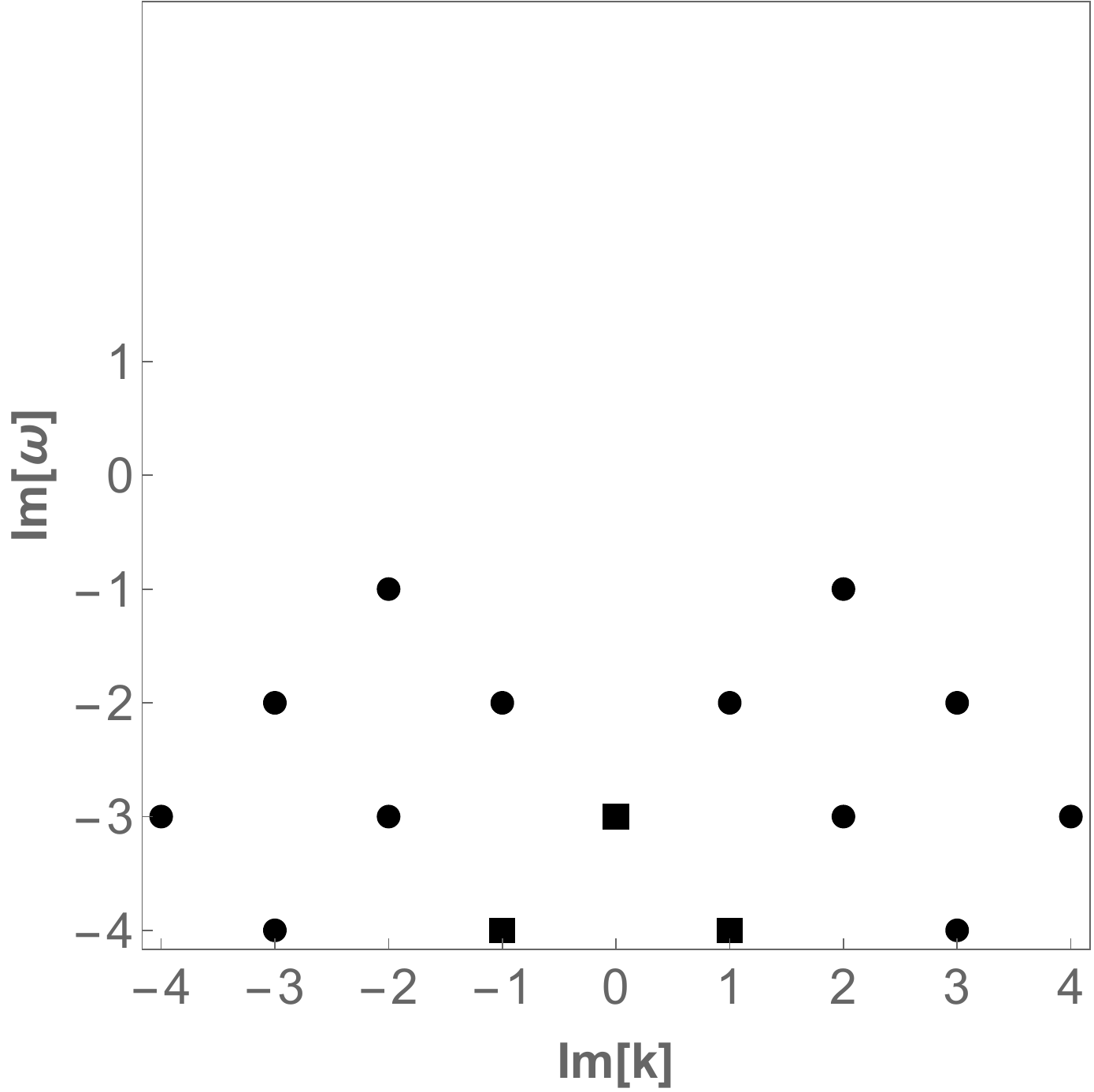}
    \label{holscalpspb}
    }
    \subfigure[$d=4, \Delta = 3.8$]{\includegraphics[width=4.8cm]{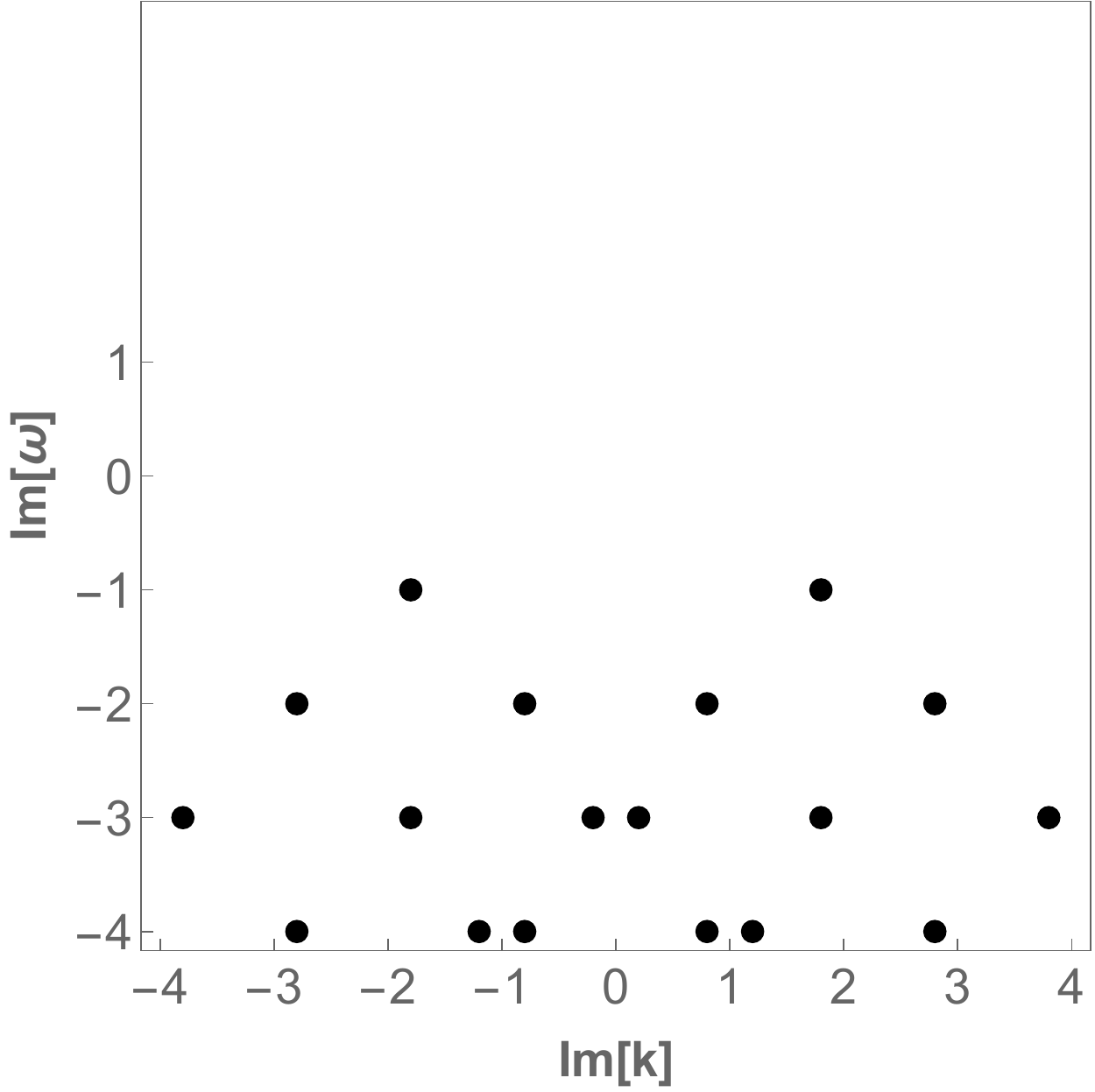}
    \label{holscalpspc}
    }
\caption{The pole-skipping points of scalar field  obtained from near horizon analysis. Black circles denote type-I while Black squares denote type-II points}
\label{holscalpsp}
\end{figure}

Figure \ref{holscalpsp} shows the types of the pole-skipping points of the scalar field case obtained by the near horizon analysis. The black circles represent the type-I points and the black squares represent the type-II points. The corresponding results obtained by the Green's function computations are shown in 
 Figs.~\ref{fig:regscalar4.2}, \ref{fig:regscalar4}, \ref{fig:regscalar3.8}, where the type-I points and type-II points are marked as white circles and squares. We find that two plots agree with each other. 

 
\begin{figure}
 \centering
    \subfigure[$d=4, \Delta=3$]{\includegraphics[width=4.8cm]{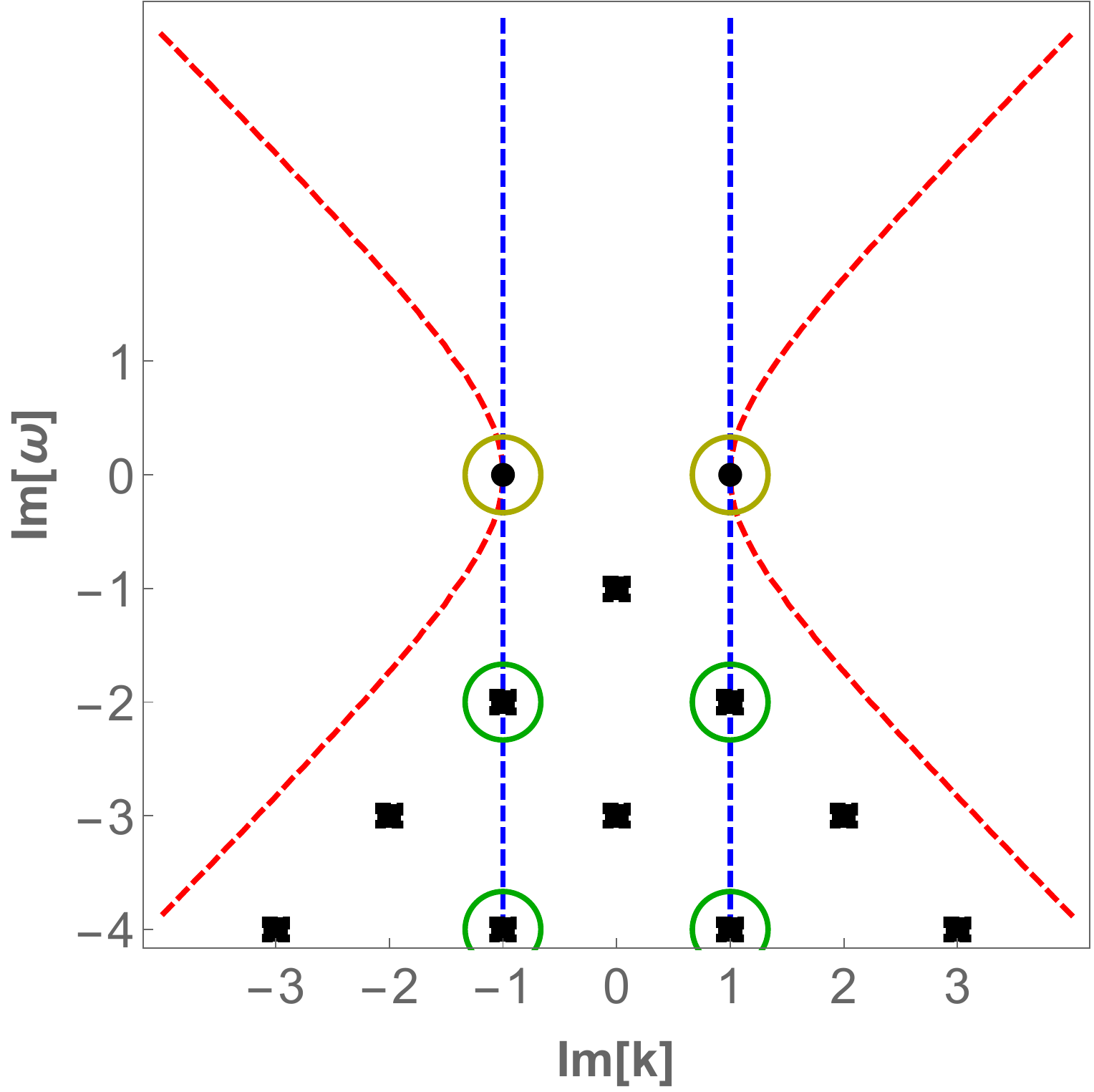}
    \label{diffpspsa}}
    \subfigure[$d=5, \Delta=4 $]{\includegraphics[width=4.8cm]{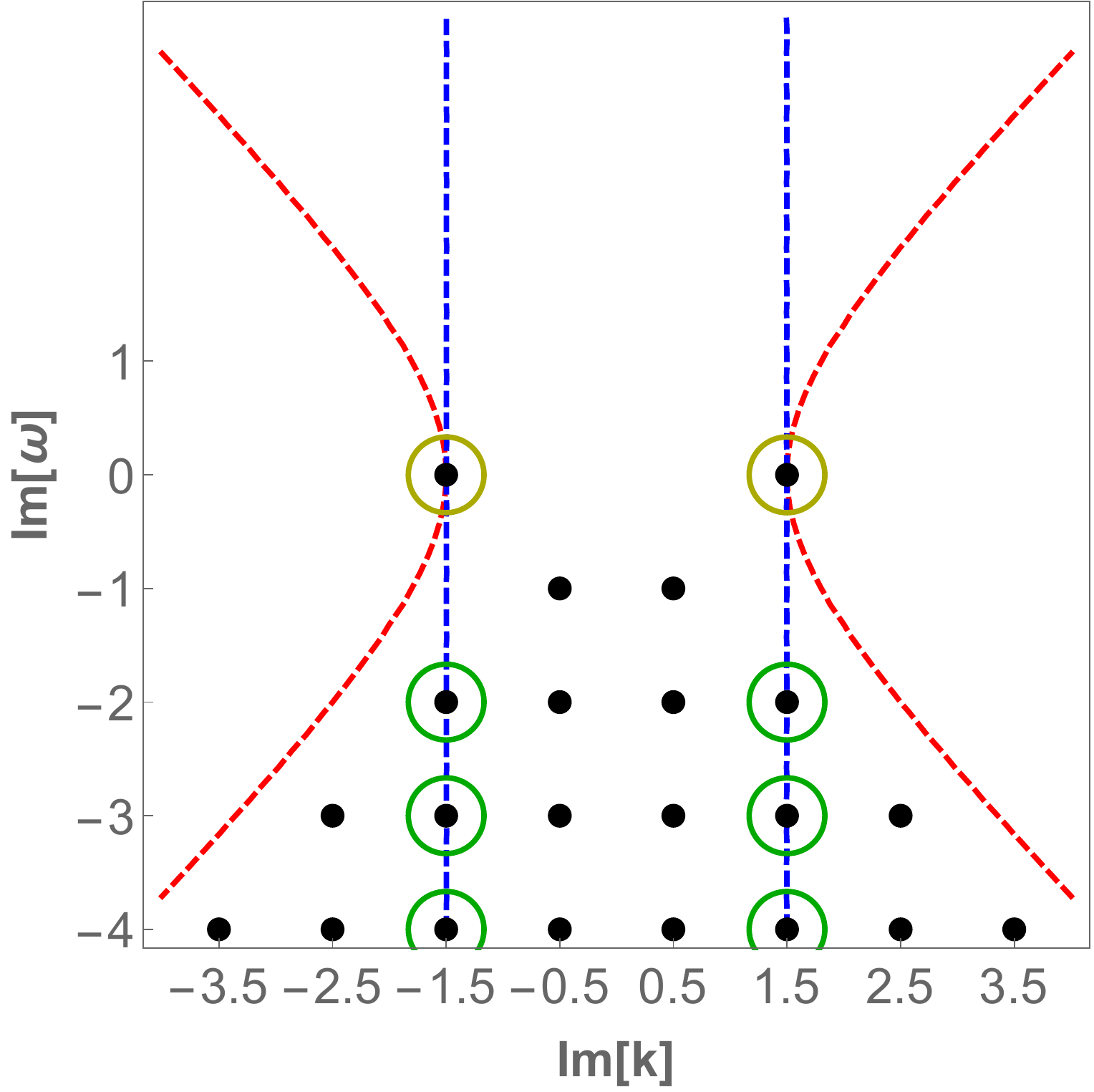}
    \label{diffpspsb}}
    \subfigure[$d=6, \Delta=5$]{\includegraphics[width=4.8cm]{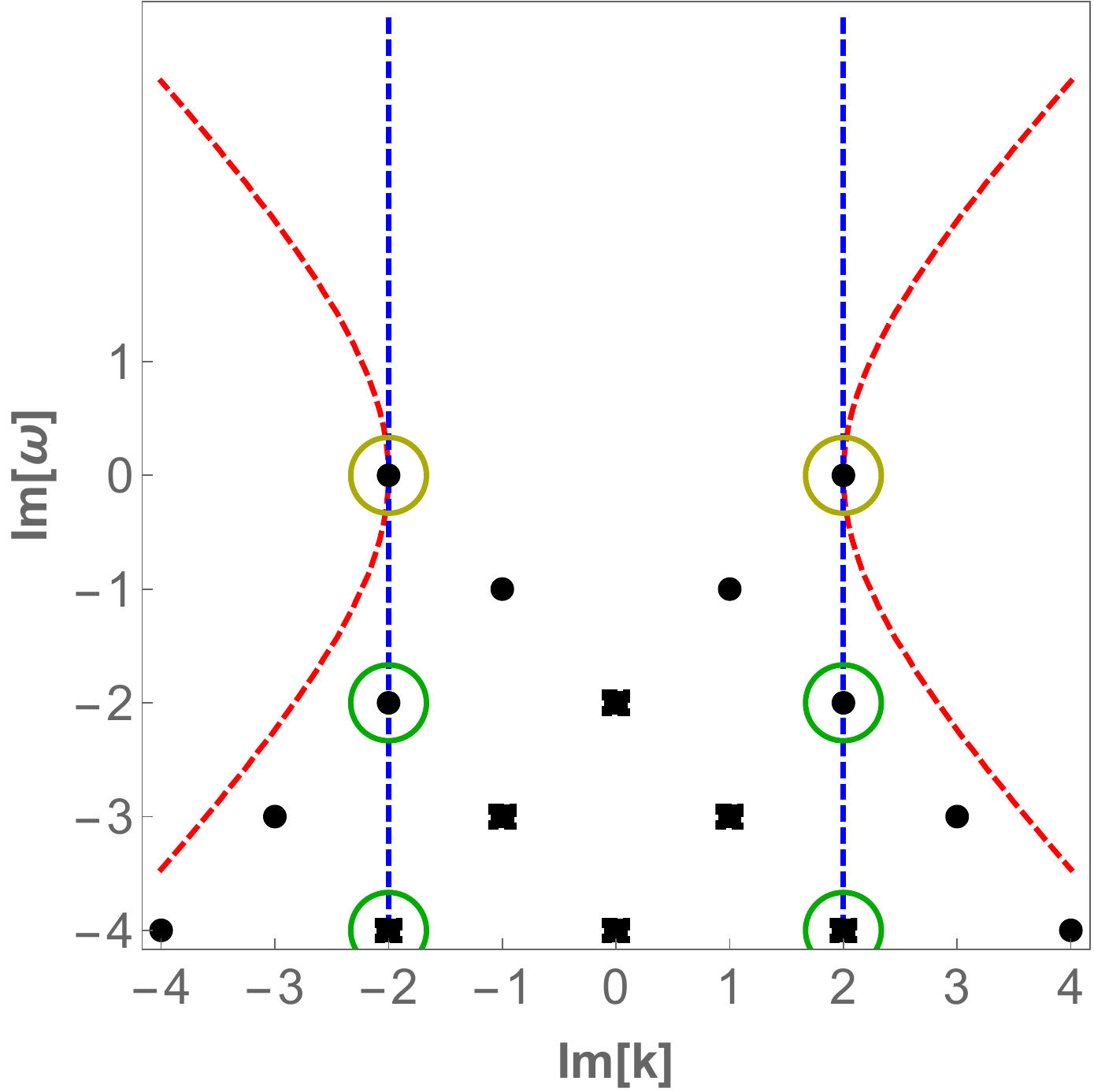}
    \label{diffpspsc}}
    \caption{The pole-skipping points of massless vector field (longitudinal channel) obtained from near horizon analysis. Black circles denote type-I while Black squares denote type-II points. Red dashed curves are lines of pole and blue curves are lines of zeroes coming from the prefactor. }
    \label{diffpsps11}
\end{figure}

Fig.~\ref{diffpsps11} shows the types of pole-skipping points of the massless vector field (longitudinal channel) case obtained by the near horizon analysis. The corresponding results obtained by the Green's function computations are shown in 
in Fig.~\ref{massless123}. The locations of pole-skipping point are the same. However, there are some differences for the types of pole-skipping points. The types of the points in the green and yellow circles in Fig.~\ref{diffpsps11} do not agree with the corresponding ones in Fig.~\ref{massless123}.

\paragraph{Prefactor issue}
The reason for this discrepancy is the following. 
In the field theory calculation, we compute the whole Green's function $G^R_{t,t}$ but in the near horizon analysis of the holographic method, we compute the pole-skipping points of a gauge invariant variable $\mathcal{U}_v$, which is related to a part ($\Pi_L$) of the retarded Green's function as follows\footnote{The procedure to obtain \eqref{retGandgageinv} is similar to the flat case shown in ~\cite{Kovtun:2005ev}}.
\begin{equation}
\label{retGandgageinv}
    G^R_{t,t} = \frac{{k}^2+\left(\frac{d-2}{2}\right)^2}{{k}^2+\left(\frac{d-2}{2}\right)^2-\omega^2}\Pi_L, 
    \qquad \Pi_L = \frac{\mathcal{U}_t^{(d-2)}}{\mathcal{U}_t^{(0)}} = \frac{\mathcal{U}_v^{(d-2)}}{\mathcal{U}_v^{(0)}} + \cdots\,,
\end{equation}
where $\mathcal{U}_t$ is a gauge invariant variable that in Schwarzschild coordinates $(t,z,\chi,\Omega_{d-2})$ can be written as
\begin{equation}
\label{rel:UtandUv}
\mathcal{U}_t(z) = e^{i \omega z_*}\mathcal{U}_v(z)=\mathcal{U}_t^{(0)} + \dots + \mathcal{U}_t^{(d-2)}z^{d-2} + \cdots \,,
\end{equation}
where $z_*$ is defined in \eqref{inEFco}. 

Note that the full Green's function $G_{t,t}^R$ consists of a prefactor and $\Pi_L$. The numerator (denominator) of this prefactor gives a extra zero (pole) condition in addition to the ones from $\Pi_L$. Thus, we have to take care of them in the near horizon analysis of $\Pi_L$ in Fig.~\ref{diffpsps11}, where the blue line is a line of zeros and the red curve is a curve of poles. 
Due to this new line of zeros coming from the prefactor, the type of pole-skipping points in the green circles must be changed (type I $\leftrightarrow$ type II). Recall that the number of lines determine the type of pole-skipping points as explained in section~\ref{sec:GFOG}. Thus, the two results in Fig.~\ref{massless123} and Fig.~\ref{diffpsps11} indeed agree. However, the ones in the yellow circles have some subtle aspects, because they are accompanied also by another red curve in Fig.~\ref{diffpsps11}. We explain this subtlety in more detail below.

\paragraph{Puzzle and Type-III pole-skipping points} 
The red curve in Fig.~\ref{diffpsps11} from the prefactor may cause a problem because this curve of poles does not exist in Fig.~\ref{massless123}. This new curve of poles, if existed, would have predicted  more pole skipping points on the curve at the {\it non-integer} $i\omega$. 

\begin{figure}
    \centering
    \subfigure[ $\Pi_L$]{\includegraphics[width=4.8cm]{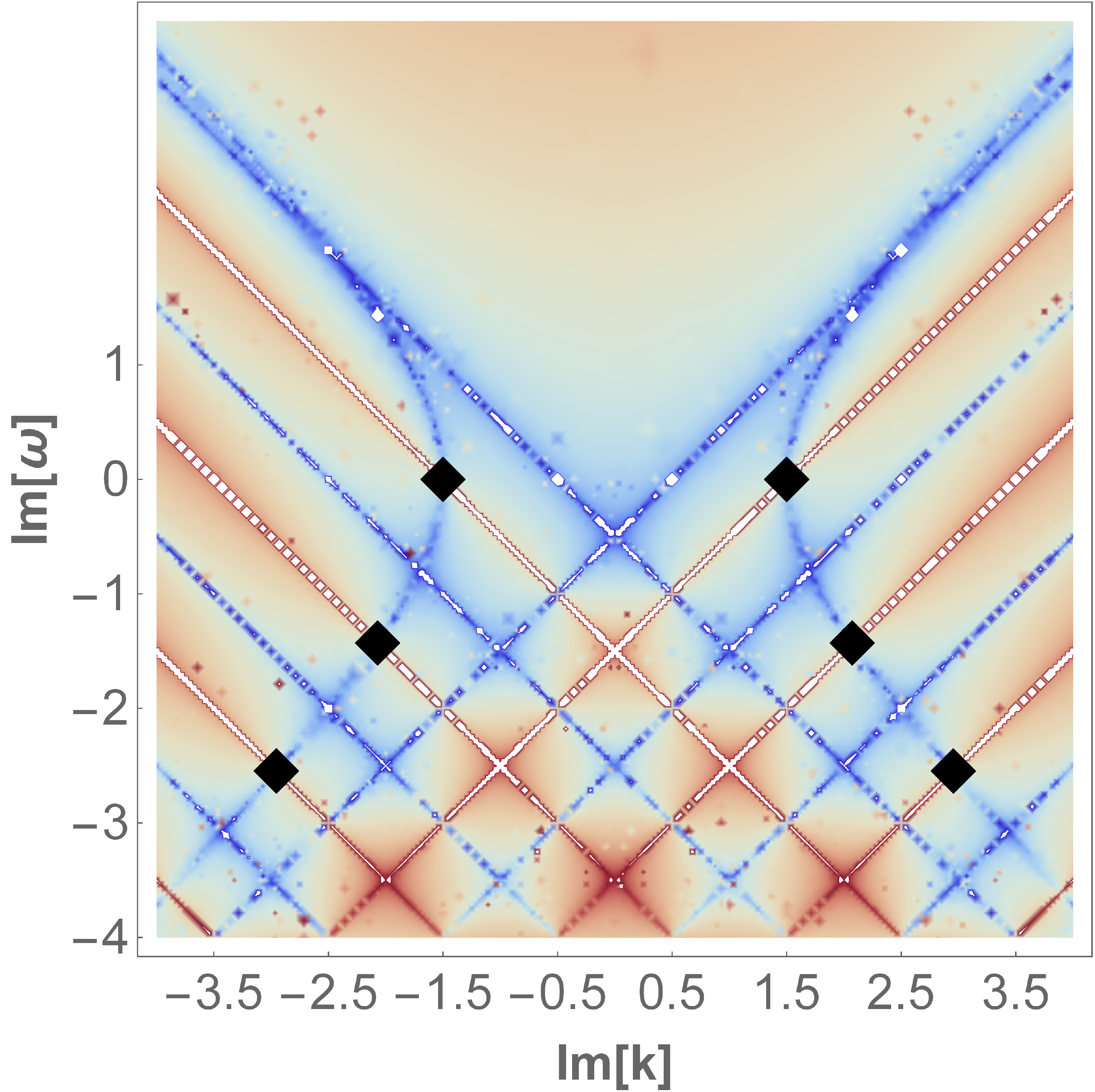}
    \label{piL}}
    \subfigure[ $G^R_{t,t}$]{\includegraphics[width=4.8cm]{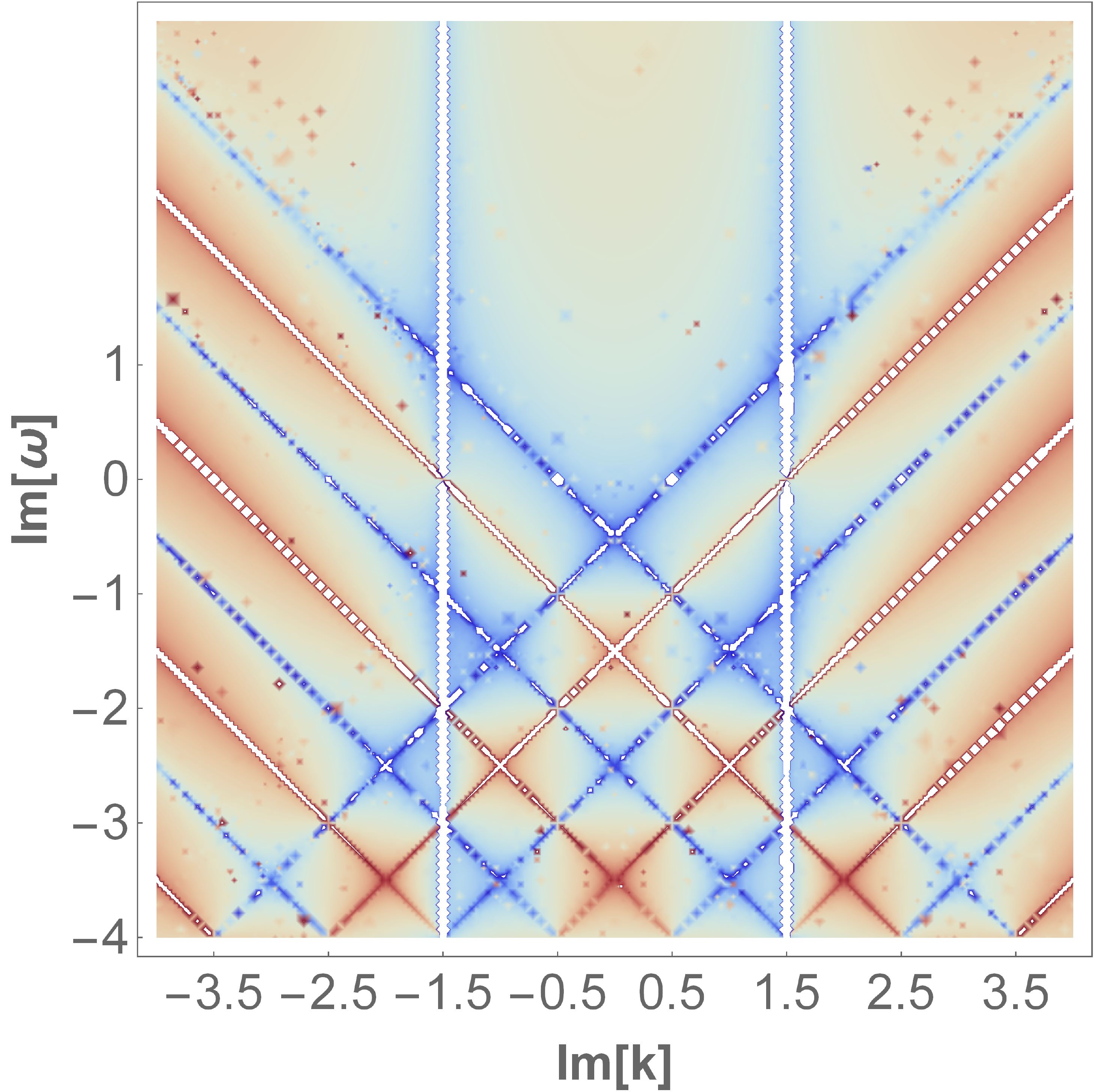}
    \label{diffGR}}
    \caption{Numerical results by holography. $\Pi_L$ and $G^R_{t,t}$ in Eq.~\eqref{retGandgageinv} when $d=5, \Delta = 4$. The right panel (b) agrees with the filed theory result, Fig.~\ref{massless123b}.}
    \label{fig:gaugeinvGR}
\end{figure}

Indeed, it turns out that we do not need to worry about it, because the red curve of poles is canceled out by the yet unrevealed zero curve in $\Pi_L$. First, let us show the existence of the curve of zeros by plotting $\Pi_L$  in Fig.~\ref{piL}, which is computed numerically by an holographic computation. Next, the full Green's function with the prefactor $G^R_{t,t}$ is shown in Fig.~\ref{diffGR}, obtained again in holography by a numerical computation. Fig.~\ref{diffGR} confirms our field theory result Fig.~\ref{massless123b} and proves that the curve of poles is precisely canceled by the curve of zeros. 

Interestingly, the existence of this curve of zeros suggests the possibility of a new type of pole-skipping point located at  non-integer values of $i\omega$, which are marked as black diamonds in Fig.~\ref{piL}.  In this case, they are removed by the prefactor, but in general this might not be the case. Let us call them type-III pole-skipping points in the case they remain as pole-skipping points.

What is the origin of the type-III points? It comes from a UV condition. On the curve $\omega^2 = k^2 + \left(\frac{d-2}{2}\right)^2$, the near boundary expansion of $\mathcal{U}_t$ is~\footnote{We do not display some terms which are irrelevant to our interest. There might be a $\log$ solution or an independent series solution starting from $z^d$ in \eqref{bdasym:Ut}. The boundary behavior of $\mathcal{U}_t$ and $\mathcal{U}_v$ are related by~\eqref{rel:UtandUv}.}
\begin{equation}
\label{bdasym:Ut}
    \mathcal{U}_t(z) = \mathcal{U}_t^{\left(0\right)} + \cdots +  0\cdot z^{d-2} + \cdots  
\end{equation}
so in this case the coefficient of $z^{d-2}$ is always zero, yielding a curve of zeros. This specific UV constraint makes pole-skipping possible when $\mathcal{U}_t^{\left(0\right)}$ becomes zero. In this case, the pole-skipping points can occur even when $i\omega$ takes {\it non-integer} values.

To discuss the type-III pole-skipping points of $\Pi_L$ in the near horizon analysis more, let us focus on the equation \eqref{diffeq} of $\mathcal{U}_v$. The coefficients $\Omega(\omega, k)$ and $\Xi(\omega, k)$ include 
\begin{align}
\frac{1}{\omega^2-\left(k^2+\left(\frac{d-2}{2}\right)^2\right)f(z)},\label{function1}
\end{align}
where $f(0)=1$ at UV, and $f(1)=0$ at IR. On the curve $\omega^2 = k^2 + \left(\frac{d-2}{2}\right)^2$, (\ref{function1}) diverges at UV $(z\sim0)$. This divergence  is related to the zero curve in $\Pi_L$. However, at IR $(z\sim1)$, (\ref{function1}) is finite except for $\omega=0$\footnote{The case $\omega=0$ is considered separately in \eqref{eq:vectorzeroomega}.}. This observation implies that the pole-skipping points of $\Pi_L$ on the zero curve $\omega^2 = k^2 + \left(\frac{d-2}{2}\right)^2$ are not captured from the near horizon analysis.  Therefore, type-III pole-skipping points  can be defined as the pole-skipping points which are not captured by the near horizon analysis.

\section{Conclusions\label{discussion}}

In this paper, we clarified general mathematical and physical properties of pole-skipping points. To achieve this goal, we take advantage of i) scalar and vector fields ii) in hyperbolic space. Regarding i) we have the following reason. Even though the original motivation of the pole-skipping point came from the metric field which is related with the energy density two point Green's function, the general properties of the pole-skipping points can be more easily understood by scalar and vector fields, partly because they have less components than the metric. Regarding ii), the hyperbolic space allows us to obtain an analytic expression for the thermal Green's function by a conformal transformation of a vacuum flat space result. In flat space with $d-1\ge2$, it is not easy to obtain an explicit analytic expression of the thermal Green's function, and one usually resort to numerics. Thanks to the analytic formulas for the Green's function, we can clearly understand the general properties of the pole-skipping points and its relation to the holographic near horizon analysis.

We classify the pole-skipping points of the Green's function in three types. 
\begin{itemize}
\item Type-I and II: The pole-skipping points arise at integer (half integer) values of $i\omega$ for bosons (fermions)\footnote{See \cite{Ceplak:2019ymw} for fermionic cases.}. At the pole-skipping points, the Green's function is not unique and depends on the path we use to approach the points in momentum space. For type-I points, the path along which the Green's function is constant, is linear, while for type-II points, it is not linear.
\item Another convenient criterion is as follows. Type I points occur when the same number of zero-lines and pole-lines intersect at a point. For type II points, a different number of zero-lines and pole-lines intersect at a point\footnote{There is a rare exception when the zero curve and pole curve are tangential to each other. It is explained around~\eqref{cgshe}.}
\item Type-III: The pole-skipping points arise at non-integer values of $i\omega$. 
\end{itemize}
From a holographic perspective, we may understand the three types in terms of a near horizon analysis.
\begin{itemize}
\item At the type-I and II pole-skipping points in the near horizon analysis, the regular incoming boundary condition is ambiguous so the Green's function at the pole-skipping points can take  any value depending on the choice of incoming boundary condition. The regularity constraints the value of $i\omega$, which takes integer values for bosons and half integer values for fermions~\cite{Blake:2019otz, Natsuume:2019vcv,Ceplak:2019ymw}
\item By the near horizon analysis, we constrain the general form of the Green's function near the pole-skipping points. See \eqref{jshg7} and \eqref{jshg74}.
\item For a massless vector field (or higher spin case), the Green's function is expressed as a prefactor times $\Pi_L$~\eqref{retGandgageinv}. In the holographic computation, we use $\Pi_L$ so the analysis alone may not be enough to detect the pole-skipping point and/or its type. Without a prefactor, as in the scalar case, if $\partial_k\det \mathcal{M}^{(n)}|_* \ne 0$, where  $\mathcal{M}^{(n)}$ is defined in \eqref{pspeom1} and its specific form is given in \eqref{pspeom123}, the pole-skipping point is of type I~\cite{Blake:2019otz, Natsuume:2019vcv}. Otherwise, it is of type-II. However, if there is a prefactor, this criteria should be modified. Some examples are shown in Sec.~\ref{Sec5}. In principle, the would-be pole-skipping point by the near horizon analysis may be removed by the prefactor. 
\item Type-III:  At the type-III pole-skipping points, contrary to type-I and type-II in the near horizon analysis, the incoming boundary condition is unique.
Even with an unique incoming boundary condition, the Green's function at the pole-skipping points can not be defined uniquely because it  still takes the form $0/0$. The value of $i\omega$ can be a non-integer because there is no constraint from the regularity condition. This type can occur because of the specific UV structure of the theory, not only because of the IR structure.
\end{itemize}

The type-II pole-skipping point was originally observed in~\cite{Blake:2019otz}, where it was called an `anomalous point'.
We choose to call it type-II pole-skipping point because `anomalous' gives an impression that something strange happens.
In our opinion, the type-II point is one natural possibility for pole-skipping.\footnote{It was first noted in \cite{Natsuume:2019xcy, Natsuume:2019vcv}.}
From a mathematical point of view the  Green's function {\it near} the pole-skipping point is always well defined as pointed out in~\cite{Blake:2019otz}, and the difference between type-I and type-II points is simply the path on which the Green's function is constant: is the path linear, quadratic or some higher order polynomial? In other words, the difference between type-I and type-II points is how to define the limiting value of the Green's function when it approaches to the pole-skipping point. From a physical point of view, these path-dependent definitions of the limiting value may be meaningful, when they give different physical interpretation. For the type-II case, the path is not necessarily linear.

The type-III point is interesting because in this case, $i\omega$ can take non-integer values. To understand this new possibility, it is helpful to revisit the definition of the pole-skipping point from holographic Green's function perspective. Let us consider the UV expansion of the field $\Phi$
\begin{equation}
\label{bdasym:Ut11}
    \Phi(z) = \mathcal{A}(\omega, k) + \cdots +  \mathcal{B}(\omega, k) z^{\#} + \cdots \,,  
\end{equation}
where $\#$ is a positive integer for a normalizable mode. The very definition of the pole-skipping point is the point where the Green's function is not well-defined.  A possible way to identify pole-skipping points is to find specific values of frequency and wave-number $(\omega_*, k_*)$ where  $\mathcal{A}(\omega_*, k_*)=\mathcal{B}(\omega_*, k_*)=0$.
 Whatever mechanism we have, once we find $(\omega_*, k_*)$ such that $\mathcal{A}(\omega_*, k_*)=\mathcal{B}(\omega_*, k_*)=0$, we have a pole-skipping point.  As in the massless vector case, a specific UV condition may give a zero condition such that $\mathcal{B}(\omega_*, k_*)=0$, which defines a zero-curve in $(\omega, k)$ space. If a lines of poles meets this zero-curve, then we have a pole-skipping point, when $\omega$ can naturally be a non-integer. 

One might wonder what happens to the importance of the ambiguity of the incoming boundary condition to have a non-unique Green's function at the pole-skipping point. We agree that this condition is very important, but it is one useful criterion to find a pole-skipping point but not a necessary condition. 

In principle, the non-uniqueness of the Green's function is not equivalent to $0/0$ itself.  
If $i\omega$ is a positive integer, we find that the solution has a free parameter, say $\alpha$, from the near horizon expansion: 
\begin{equation}
    \Phi(z) = 1 + \cdots +  \alpha (1-z)^{i\omega} + \cdots \,,  
\end{equation}
Thus, the Green's function is a function of $\alpha$ 
\begin{equation}
    G^R(\alpha) \sim \frac{\mathcal{A}(\omega_*, k_*;\alpha)}{\mathcal{B}(\omega_*, k_*;\alpha)} \,,
\end{equation}
which means $G^R(\alpha)$ can take any value depending on $\alpha$. Thus it implies the non-uniqueness of the Green`s function but, strictly speaking, it does not imply that it takes the form $0/0$.  In short, the requirement of two independent incoming boundary conditions at the horizon is a very convenient technique with beautiful physics insight to find a Green's function of the form $0/0$, but in principle, it is neither a sufficient nor a necessary condition.

\begin{figure}
\centering
    \includegraphics[width=5.5cm]{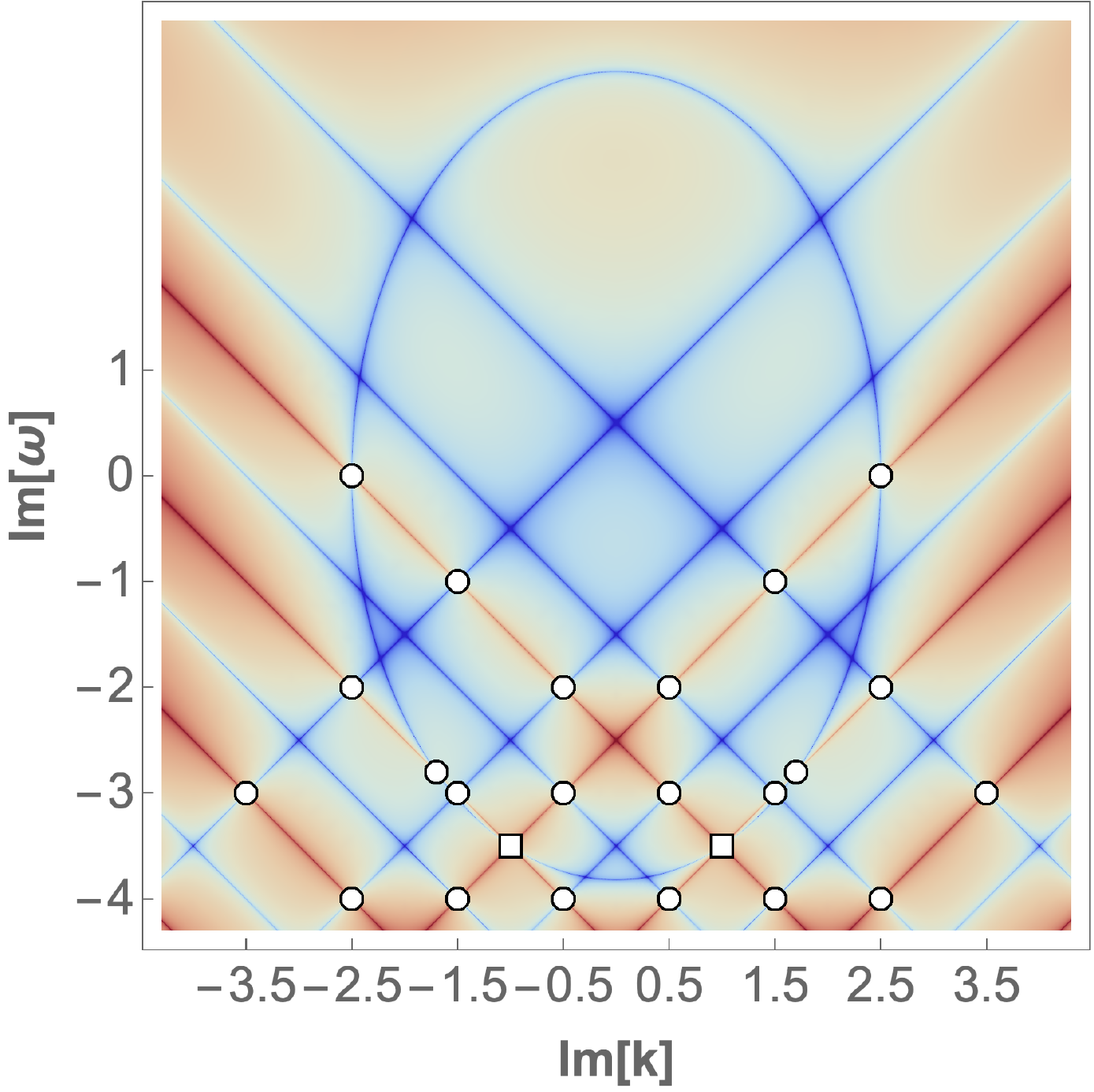}\label{fig:vector1}
    \caption{$\log |\mathcal{G}_V^{\Delta}(\mathrm{Im}[\omega],\mathrm{Im}[k])|$ at $d=4,\,\Delta=4.5\,(\delta=2.5)$. The blue lines and red lines represent zeros and poles of the Green's function, respectively. The white circles and squares are type I and type II pole-skipping points respectively. Compared to the massless $(\Delta=d-1)$ vector's Green's function, massive $(\Delta\neq d-1)$ vector's one have elliptic blue line instead of the vertical blue lines in Fig.\ref{massless123}. In general, such zero line's shape is conical section. Thus, its shape can also be circle, parabola and hyperbola depending on the value of $\Delta, d$. Note that there exist the pole-skipping points with non-integer value of $\mathrm{Im}[\omega]$ around $-2.8,-3.5$ for each.} 
    \label{confpolesmry}
\end{figure}

In our paper, the type-III point was not observed in the Green's function but rather in the `gauge-invariant' Green's function. This would-be type-III point in the massless vector case turned out removed by a prefactor. This cancellation may indicate that the type-III point is possibly an artifact of the variable choice. However, we think it is still possible to obtain a type-III point in some Green's function by the mechanism we explained above, by the UV condition.  
One potential example may be the massive vector case. In the Green's function for a massive vector field, the pole-skipping points with non-integer $i\omega$ were observed \cite{Ahn:2020bks} as seen in Fig.~\ref{confpolesmry}. See the white circles around Im[$\omega$]$=-2.8,-3.5$. It would be interesting to analyse this case in more detail from the holographic perspective. 
Another possible example is the Green's function of energy density in the large $q$ limit of SYK chain \cite{Choi:2020tdj}, where the non-integer $i\omega$ pole-skipping point was discovered. 
We stress again that the important feature of the type-III point is it naturally explains non-integer $i\omega$.
  We leave further analysis of the type-III point as future work. Last but not least, it will be interesting to figure out some phenomenological effects of the pole-skipping points and their types we revealed in this paper.





\acknowledgments
We would like to thank Richard A.~ Davison and Ioannis Papadimitriou for
valuable discussions and comments.
This work was supported by Basic Science Research Program through the National Research Foundation of Korea(NRF) funded by the Ministry of Science, ICT \& Future
Planning (NRF-2017R1A2B4004810) and the GIST Research Institute(GRI) grant funded by the GIST in 2020. 
H.-S. Jeong was supported by Basic Science Research Program through the National Research Foundation of Korea(NRF) funded by the Ministry of Education(NRF-2020R1A6A1A03047877).
V. Jahnke was supported by Basic Science Research Program through the National Research Foundation of Korea(NRF) funded by the Ministry of Education(NRF-2020R1I1A1A01073135). 
M. Nishida was supported by Basic Science Research Program through the National Research Foundation of Korea(NRF) funded by the Ministry of Education(NRF-2020R1I1A1A01072726).
We also would like to
thank the APCTP(Asia-Pacific Center for Theoretical Physics) focus program, ``Quantum Matter from the Entanglement and Holography" in Pohang, Korea for the hospitality during our visit, where part of this work was done.

\appendix 


\providecommand{\href}[2]{#2}\begingroup\raggedright\endgroup
\bibliographystyle{JHEP}

\end{document}